\documentclass[10pt,twocolumn,showpacs,amsmath,amssymb, aps, prb]{revtex4-1}

\usepackage{graphicx}
\usepackage{dcolumn}
\usepackage{bm}
\usepackage{hyperref}
\usepackage[cp1250]{inputenc}
\usepackage[usenames,dvipsnames]{color}
\usepackage{tabularx}

\newcommand{\veck}{{\mathbf k}}

\newcommand{\vecS}{{\mathbf S}}
\newcommand{\vecr}{{\mathbf r}}
\newcommand{\vecR}{{\mathbf R}}
\renewcommand{\today}{Resubmitted version of April 9, 2016}

 \definecolor{darkgreen}{rgb}{0,0.5,0} %
 \usepackage{color}
 \usepackage[normalem]{ulem}
 \usepackage{cancel}
             
\begin{document}

\title{Quasi-particle bands and structural phase transition of iron\\
from Gutzwiller Density-Functional Theory}

\author{Tobias Schickling}
\author{J\"org B\"unemann}
\author{Florian Gebhard}
\email{florian.gebhard@physik.uni-marburg.de}
\affiliation{Fachbereich Physik, Philipps-Universit\"at Marburg,
D-35032 Marburg, Germany}
\author{Lilia Boeri}
\affiliation{Institute for Theoretical and Computational Physics, 
TU Graz, A-8010 Graz, Austria}

\date{\today}

\begin{abstract}%
We use the Gutzwiller Density Functional Theory to calculate
ground-state properties and bandstructures of iron 
in its body-centered-cubic (bcc) and
hexagonal-close-packed (hcp) phases. 
For a Hubbard interaction~$U=9\, {\rm eV}$ and Hund's-rule 
coupling~$J=0.54\, {\rm eV}$ we reproduce the lattice parameter,
magnetic moment, and bulk modulus of bcc iron. For these parameters, bcc 
is the ground-state lattice structure at ambient pressure
up to a pressure of $p_{\rm c}=41\, {\rm GPa}$
where a transition to the non-magnetic hcp structure is predicted,
in qualitative agreement with experiment
($p_{\rm c}^{\rm exp}=10\ldots 15\, {\rm GPa}$). The calculated bandstructure
for bcc iron is in good agreement with ARPES measurements.
The agreement improves when
we perturbatively include the spin-orbit coupling.
\end{abstract}



\maketitle

\section{Introduction}
The theoretical description of the structural and electronic properties of
iron poses an interesting but difficult problem.
Since iron is an essential element in the inner core of the earth
it is desirable to know its phase diagram 
over a wide temperature and pressure range.
However, basic calculations at ambient pressure and zero temperature
reveal the intricacy of the iron problem.
Ab-initio calculations with Density Functional Theory (DFT)
in the Local Density Approximation, DFT(LDA), 
predict a wrong lattice structure for the ground state, namely
face-centered-cubic (fcc) 
or hexagonal-close-packed (hcp).~\cite{PhysRevLett.54.1852,PhysRevB.67.214518}
Employing DFT with a Generalized Gradient Approximation, DFT(GGA),
standard bandstructure theory recovers the experimentally observed
ferromagnetic body-centered-cubic (bcc) structure 
and the bcc-hcp transition.~\cite{PhysRevB.50.6442,PhysRevB.58.5296} 
Using the standard GGA functional of Perdew, Burke, and Ern\-zerhof (PBE),~\cite{PBE}
a good description of the lattice parameter, magnetization, and compressibility
of iron, cobalt, and nickel is obtained.~\cite{PhysRevB.60.791,Taiwanesen}
However, as for nickel,~\cite{1367-2630-16-9-093034}
both DFT functionals lead to a $d$-electron bandwidth that is too large
and quasi-particle masses at the Fermi edge that 
are too small in comparison with experiment.~\cite{PhysRevLett.110.117206}
This indicates that neither of the DFT functionals takes 
into account the correlations among the $3d$ electrons in an optimal way.

Coulomb correlations for electrons in narrow bands are often modeled
by a purely local, Hubbard-type 
interaction.~\cite{Hubbard1963,Gutzwiller1963,Hubbard1964,Gutzwiller1964}
Typically, the effective atomic interactions are parameterized by 
an intra-orbital Hubbard interaction~$U$ and
a Hund's-rule coupling~$J$.
Unfortunately, Hubbard models pose a notoriously difficult
many-body problem. 

Over the past three decades, two many-body approaches
emerged that permit a treatment
of Hubbard-type interactions in the limit of infinite lattice
coordination number in combination with DFT.
First, the Dynamical Mean-Field Theory 
(DMFT)~\cite{RevModPhys.78.865,VollhardtAnnalen}
maps the problem onto a single-impurity model.
The spectral function of the (multi-orbital) single-impurity model is 
calculated numerically,
typically using Quantum Monte-Carlo.~\cite{RevModPhys.83.349}
With DMFT(QMC), 
iron's structural and magnetic transitions
at temperatures $T>1000\, {\rm K}$ were studied in
Ref.~[\onlinecite{0953-8984-26-37-375601}]
(and references therein), and the equation-of-state for the bcc-hcp
phase transition was investigated in Ref.~[\onlinecite{PhysRevB.90.155120}].
Second, the Gutzwiller many-body wavefunction is combined with  
DFT(LDA) to treat local interactions variationally (`LDA+Gutzwiller',
`Gutzwiller-DFT'), see Refs.~[\onlinecite{Ho2008,Deng2008,
Deng2009,Wang2008,Wang2010,
Weng2011,Yao2011a,PhysRevB.84.205124,
PhysRevB.85.035133,PhysRevB.87.045122,
PhysRevLett.111.196801,PhysRevB.90.161104,
PhysRevB.89.165122,1367-2630-16-9-093034,
PhysRevX.5.011008}]. 
Recently, the LDA+Gutz\-wil\-ler approach was applied to a number of transition metals
including nickel, iron, and iron 
pnictides.~\cite{PhysRevB.90.125102,Deng2009,Wang2010,Yao2011a,Schickling2012}

The DMFT becomes exact in infinite dimensions but
the numerical effort is quite considerable which often makes
further simplifications of the interactions advisable.~\cite{0953-8984-26-37-375601} 
Moreover, the computational demands
limit the studies to elevated temperatures and prohibits an extensive scan
in the $(U,J)$ parameter space.
On the other hand, the Gutzwiller approach is a variational method at zero temperature
and therefore best suited for the calculation of ground-state properties
such as the lattice parameter, magnetization, and bulk modulus.
Dynamical properties, however, can only be described within the
quasi-particle picture, i.e., the method provides a bandstructure
but no quasi-particle lifetimes. Since the Gutzwiller-DFT is computationally much cheaper
than DFT(LDA-DMFT), it permits a survey of the $(U,J)$ parameter space
for iron. 

In this study, we present comprehensive 
results for iron obtained from the LDA+Gutzwiller method.
We obtain the experimental values for the lattice parameter, magnetization, 
and compressibility, and provide a qualitatively correct description
of the structural transition from bcc to hcp iron under pressure.
When we take spin-orbit effects into account perturbatively,
we obtain a good agreement with ARPES measurements
of the quasi-particle bandstructure.~\cite{PhysRevB.72.155115}

Our work is structured as follows.
In Sect.~\ref{sec:method},
we summarize the Gutzwiller-DFT that was derived in detail in
Ref.~[\onlinecite{1367-2630-16-9-093034}]; here, we generalize it to the case
of more than one atom per unit cell.
In Sect.~\ref{sec:resultsgs} we discuss our results
for the ground-state properties of iron.
In Sect.~\ref{sec:bands}, we present the respective quasi-particle bandstructures
and compare them to ARPES data for bcc iron.
Conclusions, Sect.~\ref{sec:conclusions}, close our presentation.

\section{Method}
\label{sec:method}

The Gutzwiller-DFT and the minimization algorithm which we use in this 
work have been discussed in detail 
in Refs.~[\onlinecite{1367-2630-16-9-093034}]
and~[\onlinecite{Buenemann-2012-b}]. Therefore we will only 
summarize the main ideas of these methods in the present 
section and focus  on some 
particular aspects that are relevant in our calculations for iron.

\subsection{Gutzwiller-DFT}
\label{sec:methodA}

Instead of a single-particle 
reference system that leads to the standard Kohn-Sham equations,
the Gutzwiller-DFT employs a many-particle reference system. 
It explicitly takes into account 
the local Coulomb interaction (on lattice sites $\vecR$) 
\begin{eqnarray}
\hat{V}_{\rm loc} &=& \sum_{\vecR} \hat{V}_{\rm loc}(\vecR) \; , \\
\hat{V}_{\rm loc}(\vecR) 
&=&  
 \sum_{\gamma_1,\ldots,\gamma_4}
U^{\gamma_1, \gamma_2}_{\gamma_3, \gamma_4}\,
\hat{c} _{\vecR,\gamma_1}^+ \hat{c} _{\vecR,\gamma_2}^+ 
\hat{c} _{\vecR,\gamma_3}^{\vphantom{+}} \hat{c} _{\vecR,\gamma_4}^{\vphantom{+}} 
\label{eq:VlocR}
\end{eqnarray}
 in those spin-orbital states 
$\phi_{\vecR,\gamma}(\vecr)\equiv \phi_{\vecR,c,\sigma}(\vecr)$   which 
 are deemed to be strongly correlated. In our iron calculations 
 these are the $e_g$ and $t_{2g}$  orbitals of the $3d$ shell.
 The explicit form of the operator $\hat{V}_{\rm loc}(\vecR)$ for $d$-orbitals
 is given in Ref.~[\onlinecite{1367-2630-16-9-093034}], see also 
appendix~\ref{app:Coulomb}.
 As also shown in that work, one obtains the following `Hubbard 
density functional' for the  many-particle reference system 
\begin{eqnarray}
E_{\rm H}\left[\left\{ n_{\sigma}(\vecr) \right\}\right] 
&=& 
K_{\rm H}\left[\left\{ n_{\sigma}(\vecr) \right\}\right] 
+ U\left[\left\{ n_{\sigma}(\vecr) \right\}\right] \nonumber \\
&& +  V_{\rm Har}\left[\left\{ n_{\sigma}(\vecr) \right\}\right] \nonumber \\
&&+ V_{\rm loc}\left[\left\{ n_{\sigma}(\vecr) \right\}\right] -
V_{\rm dc}\left[\left\{ n_{\sigma}(\vecr) \right\}\right] \nonumber \\
&& +  E_{\rm H,xc}\left[\left\{ n_{\sigma}(\vecr) \right\}\right] \; ,
\end{eqnarray}
 where $ E_{\rm H,xc}\left[\left\{ n_{\sigma}(\vecr) \right\}\right]$ 
is the exchange-correlation functional, see Sect.~\ref{ffd},  and
\begin{eqnarray}
U\left[\left\{ n_{\sigma}(\vecr) \right\}\right] 
&=&  \sum_{\sigma}\int {\rm d}\vecr 
U(\vecr) n_{\sigma}(\vecr)  \; , \\
V_{\rm Har}\left[\left\{ n_{\sigma}(\vecr) \right\}\right] 
&=& 
\sum_{\sigma,\sigma'} \int\!\!\int {\rm d}\vecr {\rm d}\vecr' 
V(\vecr-\vecr') n_{\sigma}(\vecr) n_{\sigma'}(\vecr') \; ,\nonumber \\
&& \\
\label{554}
K_{\rm H}\left[\left\{ n_{\sigma}(\vecr) \right\}\right] 
&=&\langle \Psi_{\rm H,0}^{(n)} | \hat{H}_{\rm kin}| \Psi_{\rm H,0}^{(n)} \rangle 
 \, , 
\\\label{555}
V_{\rm loc/dc}\left[\left\{ n_{\sigma}(\vecr) \right\}\right]
&=& \langle \Psi_{\rm H,0}^{(n)} | \hat{V}_{\rm loc/dc}| \Psi_{\rm H,0}^{(n)} \rangle   \; .
\label{eq:functionsHubbard}
\label{eq:UandVH}
\end{eqnarray}
Here, we introduced the periodic potential $U(\vecr)$ of the nuclei
 and the two-particle  Coulomb interaction $V(\vecr-\vecr')$. 
The state $|\Psi_{\rm H,0}^{(n)}\rangle $ minimizes, by definition, 
  the expectation value of the 
 Hamiltonian 
\begin{equation}
\hat{H}_{\rm H}=\hat{H}_{\rm kin} + \hat{V}_{\rm loc}-\hat{V}_{\rm dc}\; ,
\label{eq:HubbardHamiltonian}
\end{equation}
 for a given (and fixed) particle density $n_{\sigma}(\vecr)$. Finally, the 
`double-counting operator'  $\hat{V}_{\rm dc}$
and the corresponding functional 
$V_{\rm dc}\left[\left\{ n_{\sigma}(\vecr) \right\}\right]$
 account for the fact that the local Coulomb interaction 
between electrons in the correlated orbitals
is already included in the Hartree energy and the 
exchange-correlation 
functional. Unfortunately, there is no systematic way to derive 
$\hat{V}_{\rm dc}$
for a given local Hamiltonian $\hat{V}_{\rm loc}$. We work
with the widely used form of the double-counting 
functional,~\cite{PhysRevB.44.943,PhysRevB.52.R5467,RevModPhys.78.865,Deng2009} 
see appendix~\ref{app:Coulomb}.

In contrast to the corresponding Kohn-Sham function\-al, 
$E_{\rm H}\left[\left\{ n_{\sigma}(\vecr) \right\}\right]$ cannot be 
 minimized without further approximations. In fact, it cannot even 
 be evaluated because $\hat{H}_{\rm H}$ is a many-particle Hamiltonian. 
 We therefore use Gutzwiller wavefunctions for the evaluation 
 of~(\ref{554}) and~(\ref{555}).  They are defined as
\begin{eqnarray}
|\Psi_{\rm G}\rangle&=& \hat{P}_{\rm G} | \psi_0 \rangle \; ,\\\label{wq}
\hat{P}_{\rm G}&=& \prod_{\vecR} \sum_{\Gamma,\Gamma'}\lambda_{\Gamma,\Gamma'}(\vecR)
\hat{m}_{\vecR;\Gamma,\Gamma'} \; ,
\end{eqnarray}
where $| \psi_0 \rangle$ is a single-particle product state,
and $\lambda_{\Gamma,\Gamma'}(\vecR)$ are the elements of the 
variational parameter matrix $\tilde{\lambda}(\vecR)$.
We further introduced the eigenstates  $|\Gamma \rangle_{\vecR}$
of $\hat{V}_{\rm loc}(\vecR)$ and the operator
\begin{equation}
\hat{m}_{\vecR;\Gamma,\Gamma'} \equiv 
|\Gamma \rangle_{\vecR} {}_{\vecR} \langle \Gamma' |
\; .
\label{eq:Vloclocal}
\end{equation}
In our calculations for iron, we work with a diagonal and lattice-site independent 
variational parameter matrix 
$\lambda_{\Gamma,\Gamma'}(\vecR)=\delta_{\Gamma,\Gamma'}\lambda_{\Gamma}$, 
i.e., our energy functional 
depends on $n_{\Gamma}=2^{10}=1024$ variational parameters $\lambda_{\Gamma}$. 
Note that non-diagonal variational parameters do not substantially change
the results in our high-symmetry cubic situation in the absence of spin-orbit coupling. 
Moreover, we do not find it necessary to implement symmetry relations among the
1024~variational parameters. This has been done in 
Ref.~[\onlinecite{PhysRevB.90.125102}] where also
non-diagonal variational parameters $\lambda_{\Gamma,\Gamma'}$ 
were taken into account.

The evaluation 
of Eqs.~(\ref{554}) and~(\ref{555}) still is a difficult many-particle problem.
It can be solved in the limit of infinite spatial dimensions 
where one obtains an analytical energy functional. 
Using this energy functional in calculations on finite-dimensional systems, 
as done in this work, 
is usually denoted as the `Gutzwiller approximation' to the energy 
functional~$E_{\rm H}\left[\left\{ n_{\sigma}(\vecr) \right\}\right]$.

We shall not repeat here the details of the Gutzwiller approximation 
or the structure of the resulting energy functional because it has been 
thoroughly discussed in earlier work. Obviously, one obtains a functional
of the form
\begin{equation}
E_{\rm H}=E_{\rm H}(| \psi_0 \rangle,\{\lambda_{\Gamma}\})
\end{equation}
that depends on the single-particle state  $| \psi_0 \rangle$
 and the  variational parameters 
$\lambda_{\Gamma}$. The minimization of the energy functional
with respect to $| \psi_0 \rangle$ leads to an 
effective  single-particle Schr\"odinger equation for 
   $| \psi_0 \rangle$. This `Gutzwiller--Kohn-Sham equation' 
is the equivalent
to the Kohn-Sham equation in ordinary DFT calculations and is solved by 
an adapted version of the open source {\sc Quantum ESPRESSO} 
code.~\cite{0953-8984-21-39-395502}
 
For the minimization with respect to the variational parameters 
$\lambda_{\Gamma}$ (`inner minimization'), 
we use an algorithm whose elements were discussed 
in Ref.~[\onlinecite{Buenemann-2012-b}]. The new feature
of our present calculations comes from the fact that in hcp iron we 
have a unit cell with two iron atoms. 
Since both iron sites have the same point symmetry one could actually use the 
existing minimization algorithm.~\cite{Buenemann-2012-b}
For later use and testing purposes, however, we developed
a code that is capable to carry out the inner minimization 
for a large number of in-equivalent correlated atoms per unit cell.   
We explain this algorithm in the following Sect.~\ref{12w}. 

\subsection{Inner minimization for systems with multiple atoms per unit cell}
\label{12w}

Let $l=1,\ldots,n_l$ be the label for the 
 $n_l$ atoms in the unit cell and $\tilde{\lambda}^l$  
 the corresponding matrices of  variational parameters. 
Then, for a translationally invariant 
 system, the energy functional of the Gutzwiller 
 approximation has the form
\begin{eqnarray}
E(\{\tilde{\lambda}^l\}) &=&\sum_{l}V_{l;{\rm loc}}(\tilde{\lambda}^l)\nonumber\\
&&+\sum_{l}\!\sum_{\gamma_1,\gamma_2}q^{l;\gamma_2}_{\gamma_1}(\tilde{\lambda}^l)
K^l_{\gamma_1,\gamma_2}
+{\rm c.c.} \label{778}\\
&&+\sum_{l,l'}\!\sum_{\gamma_1,\ldots,\gamma_4}
q^{l;\gamma_2}_{\gamma_1}(\tilde{\lambda}^l)
\left(q^{l';\gamma_4}_{\gamma_3}(\tilde{\lambda}^{l'})\right)^{*}
I^{l,l'}_{\gamma_1,\gamma_2,\gamma_3,\gamma_4}\, .\nonumber
\end{eqnarray}
The third line describes the hopping of electrons between 
correlated orbitals while the second line includes all 
contributions from hopping processes into non-correlated
orbitals. During the inner minimization, the tensors $K^l_{\gamma_1,\gamma_2}$ and 
$I^{l,l'}_{\gamma_1,\gamma_2,\gamma_3,\gamma_4}$ are 
just numbers that result from the solution of the 
Gutzwiller--Kohn-Sham equation. 
Note that in our calculations for iron, the 
renormalization matrices $q^{l;\gamma_2}_{\gamma_1}$ are real and
diagonal,  $q^{l;\gamma_2}_{\gamma_1}=\delta_{\gamma_1,\gamma_2}
q^{l;\gamma_1}_{\gamma_1}$, which simplifies the energy functional considerably.
 
The energy functional~(\ref{778}) needs to be minimized with respect 
to all matrices $\tilde{\lambda}^l$. Even with our diagonal 
Ansatz for $\tilde{\lambda}^l$, however, the total number of 
variational parameters $n_{\rm tot}=n_{\Gamma}\times n_l$ would become 
prohibitively large if we tried to minimize straightforwardly systems with different
atoms in the unit cell. Instead of minimizing~(\ref{778})  directly with respect to 
all matrices $\tilde{\lambda}^l$ simultaneously, we therefore use
the following scheme.
\begin{itemize}
\item[(i)]
Start with some initial values for the matrices $\tilde{\lambda}^l_0$
and the corresponding renormalization matrices
$q^{l;\gamma_2}_{\gamma_1;0}\equiv q^{l;\gamma_2}_{\gamma_1}(\tilde{\lambda}^l_0)$, 
e.g., the values in the non-interacting limit.
\item[(ii)]
Minimize the $n_l$ individual energy functionals
\begin{eqnarray}\nonumber
E_l(\tilde{\lambda}^l)&=&V_{l;{\rm loc}}(\tilde{\lambda}^l)\\\nonumber
&&+\sum_{\gamma_1,\gamma_2}q^{l;\gamma_2}_{\gamma_1}(\tilde{\lambda}^l)
K^l_{\gamma_1,\gamma_2}
+{\rm c.c.}\\\nonumber
&&+\frac{1}{2}\sum_{l'}\sum_{\gamma_1,\ldots,\gamma_4}
q^{l;\gamma_2}_{\gamma_1}(\tilde{\lambda}^l)
\left(q^{l';\gamma_4}_{\gamma_3;0}\right)^{*}
I^{l,l'}_{\gamma_1,\gamma_2,\gamma_3,\gamma_4}\\
\label{44}
\end{eqnarray}
with respect to $\tilde{\lambda}^l$, e.g., 
by means of the algorithm introduced in~Ref.~[\onlinecite{Buenemann-2012-b}]. 
 \item[(iii)]
If the matrices $\tilde{\lambda}^l_1$ minimize the 
$n_l$ functionals~(\ref{44}), set $\tilde{\lambda}^l_0=\tilde{\lambda}^l_1$
and go back to step~(i) until a converged solution has been reached.
\end{itemize}
In our actual calculations, the band optimization
(Gutz\-willer--Kohn-Sham equations for $|\psi_0\rangle$) and of the 
local parameters (inner minimization) are not separated. 
After an update of the matrices $\tilde{\lambda}^l$ in step~(ii),
the matrices $q^{l;\gamma_2}_{\gamma_1}$ are recalculated. 
Then, the Gutzwiller--Kohn-Sham equation is solved again to arrive at
new values for the tensors $K^l_{\gamma_1,\gamma_2}$ and
$I^{l,l'}_{\gamma_1,\gamma_2,\gamma_3,\gamma_4}$ in eq.~(\ref{44}).
Typically, we need~10 to 15~iterations of the combined cycle of
band optimization and inner minimization to reach a converged minimum.  

\subsection{Computational details}
\label{ffd}

Our work is based on the open-source plane-wave pseudopotential code
{\sc Quantum ESPRESSO}.~\cite{0953-8984-21-39-395502}
We implemented the routines necessary for the Gutzwiller-DFT, 
as described previously in Ref.~[\onlinecite{1367-2630-16-9-093034}],
see the supplemental material for further information. 
For the Gutz\-willer--Kohn-Sham 
calculations we used the LDA exchange-correlation functional 
of Perdew and Zunger for $E_{\rm H,xc}\left[\left\{ n_{\sigma}(\vecr) \right\}\right]$;
note that our calculations start from the local spin-den\-sity approximation (LSDA) 
so that we recover the results from DFT(LSDA) for $U=J=0$.
For comparison, we also performed GGA calculations based on the 
Perdew-Burke-Ernzerhof functional.
To model the electron-core interaction, we employed ultrasoft pseudo potentials 
from the standard {\sc Quantum ESPRESSO} distribution with 
non-linear core corrections.
These are Fe.pz-nd-rrkjus.UPF (LDA), and Fe.pbe-spn-rrkjus$_{\rm p}$sl.0.2.1.UPF 
(Fe.rel-pbe-spn-rrkjus$_{\rm p}$sl.0.2.1.UPF) for (scalar) relativistic GGA calculations.
The GGA pseudopotentials yield equilibrium 
lattice constants ($a_{\rm GGA}=5.39 a_{\rm B}$),
magnetic moments ($m_{\rm GGA}=2.25 \mu_B$),
and electronic structure in excellent agreement with previous full-potential 
calculations.~\cite{PhysRevB.72.155115}

In Gutzwiller-DFT, we find that a very accurate integration
over ${\bf k}$-space is needed for a proper convergence.
Therefore, integrations over the Brillouin zone
are performed using the tetrahedron method.
For bcc iron, we use at least 624 $\veck$-points in the irreducible part 
of the Brillouin zone. For non-magnetic hcp, we work with 729 inequivalent 
$\veck$-points. For all calculations a wave cutoff of $60\,\mathrm{Ry}$ was set,
and a charge-density cutoff of $600\,\mathrm{Ry}$. 

For the construction of the $3d$-orbitals we employ the
program package {\sc poormanwannier} that is part of
the standard {\sc Quantum ESPRESSO} distribution. We
generate Wannier functions using a very large energy
window of 90 eV around the Fermi energy and the resulting
$3d$ orbitals are very close to localized $3d$ orbitals.
More details on the choice of the energy window and
implementations are given in Sect.~III A in the supplemental
material.~\cite{suppmat}
In the following we denote our implementation of the LDA+Gutzwiller scheme
as `Gutzwiller-DFT'.

\section{Ground-state properties}
\label{sec:resultsgs}

\subsection{Adjustment of the Coulomb parameters}

The Gutzwiller-DFT is not a fully ab-initio method.
The Coulomb interaction between the $3d$-electrons
is parameterized by the Hubbard interaction~$U$
and the Hund's-rule coupling~$J$. In this work,
we choose to adjust these two parameters
such that the Gutzwiller-DFT reproduces
the experimental values for the lattice parameter~$a$ and
the magnetization~$m$; see our discussion Sect.~\ref{lilia}.

The DFT(LDA) predicts a fcc or hcp crystal structure
as ground state for iron 
at ambient pressure.~\cite{PhysRevLett.54.1852,PhysRevB.67.214518}
Therefore, we do not show the results for $U=J=0$ in the following.
This serious flaw of DFT(LDA) is easily overcome
with LDA+Gutzwiller, and also 
in DFT(GGA) where $a_{\rm GGA}\approx 5.36\, a_{\rm B}$ 
and $m\approx 2.2\, \mu_{\rm B}$ 
are obtained.~\cite{PhysRevB.50.6442,PhysRevB.58.5296}
For local interactions as small as
$U \geq 1.0\,\mathrm{eV}$ and $J=0.06U$,
Gutzwiller-DFT finds the experimentally observed
ferromagnetic bcc lattice structure. Therefore,
the correct ground-state structure dominates the
$(U,J)$ phase diagram, and it is straightforward
to search for the optimal Hubbard interaction~$U$ and Hund's-rule
coupling~$J$, as was done for nickel previously.~\cite{1367-2630-16-9-093034}

\begin{figure}[ht]
\includegraphics[width=0.43\textwidth]{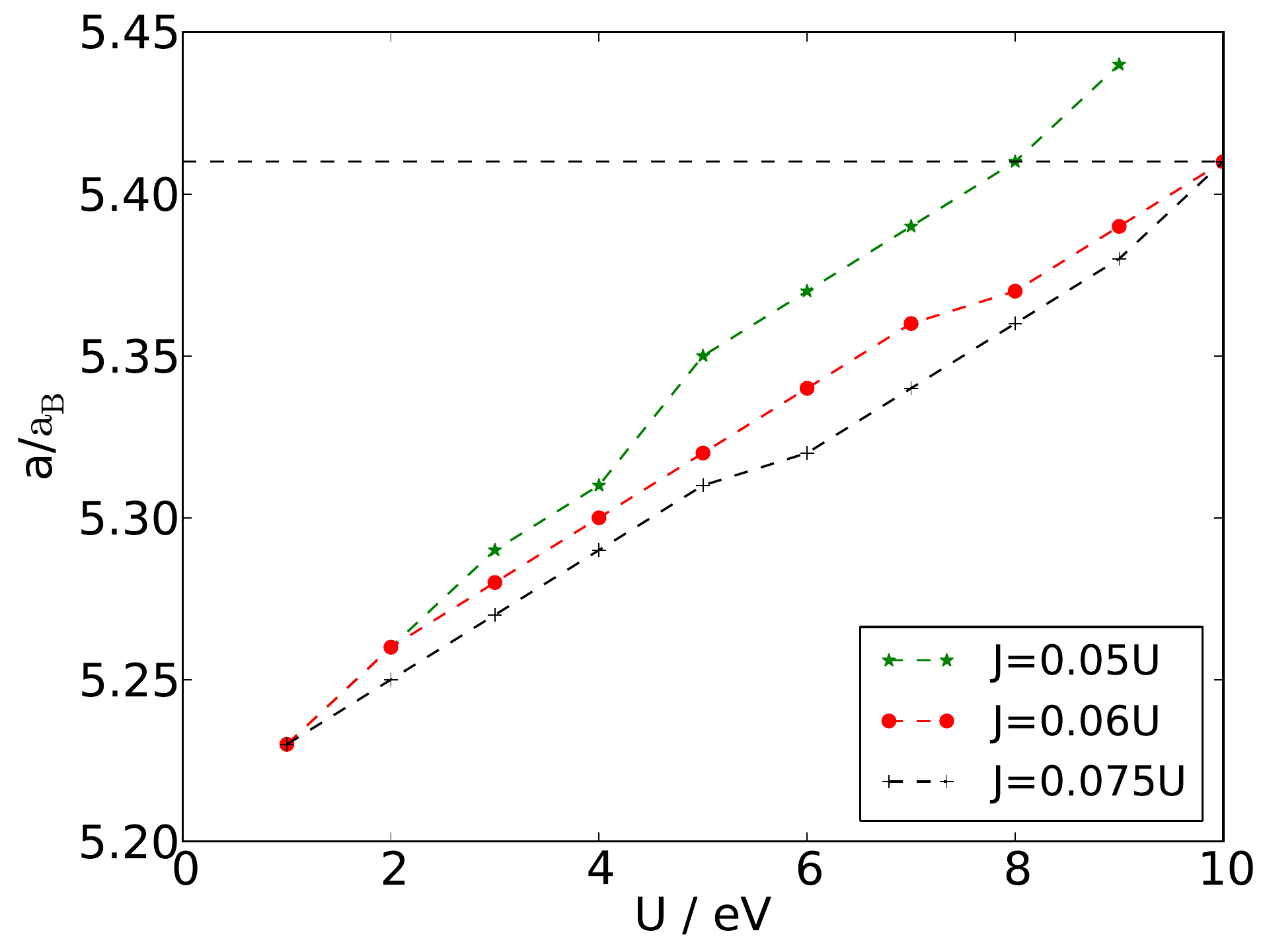}
\caption{Cubic lattice parameter~$a(U,J)$ for bcc iron in units of 
the Bohr radius~$a_{\rm B}=0.529\,\hbox{\AA}$ as a 
function of the Hubbard interaction~$U$ for $J/U=0.05, 0.06, 0.075$. 
The horizontal dashed line indicates the experimental value $a^{\rm exp}=5.42a_{\rm B}=
2.87\, \hbox{\AA}$.\label{fig:lat-const}}
\end{figure}

\subsubsection{Lattice parameter}

In Fig.~\ref{fig:lat-const} we display the bcc
lattice parameter~$a(U,J)$ as a function of~$U$ for various ratios~$J/U$. 
The horizontal dashed line indicates the experimental value,
$a^{\rm exp}=5.42a_{\rm B}=2.87\,\hbox{\AA}$.~\cite{PhysRevB.82.132409}
As also seen in nickel,~\cite{1367-2630-16-9-093034}
the lattice parameter increases monotonously as a function
of the Hubbard interaction. This effect is desired because
the DFT(LDA) considerably underestimates the lattice parameter for iron.

The influence of the Hubbard interaction is readily understood.
The Coulomb repulsion weakens the contribution of the $3d$-electrons
to the metallic binding so that the crystal is less tightly bound;
the crystal volume increases as a function of the Coulomb
repulsion. Figure~\ref{fig:lat-const} shows that
the Hund's-rule coupling~$J$ counteracts the Hubbard interaction~$U$.
The slope of $a(U,J)$ as a function of~$U$ becomes smaller
for larger~$J$. This indicates that the Hund's-rule coupling~$J$
in iron has a tendency to increase the electrons' itineracy, see below.

\begin{figure}[b]
\includegraphics[width=0.43\textwidth]{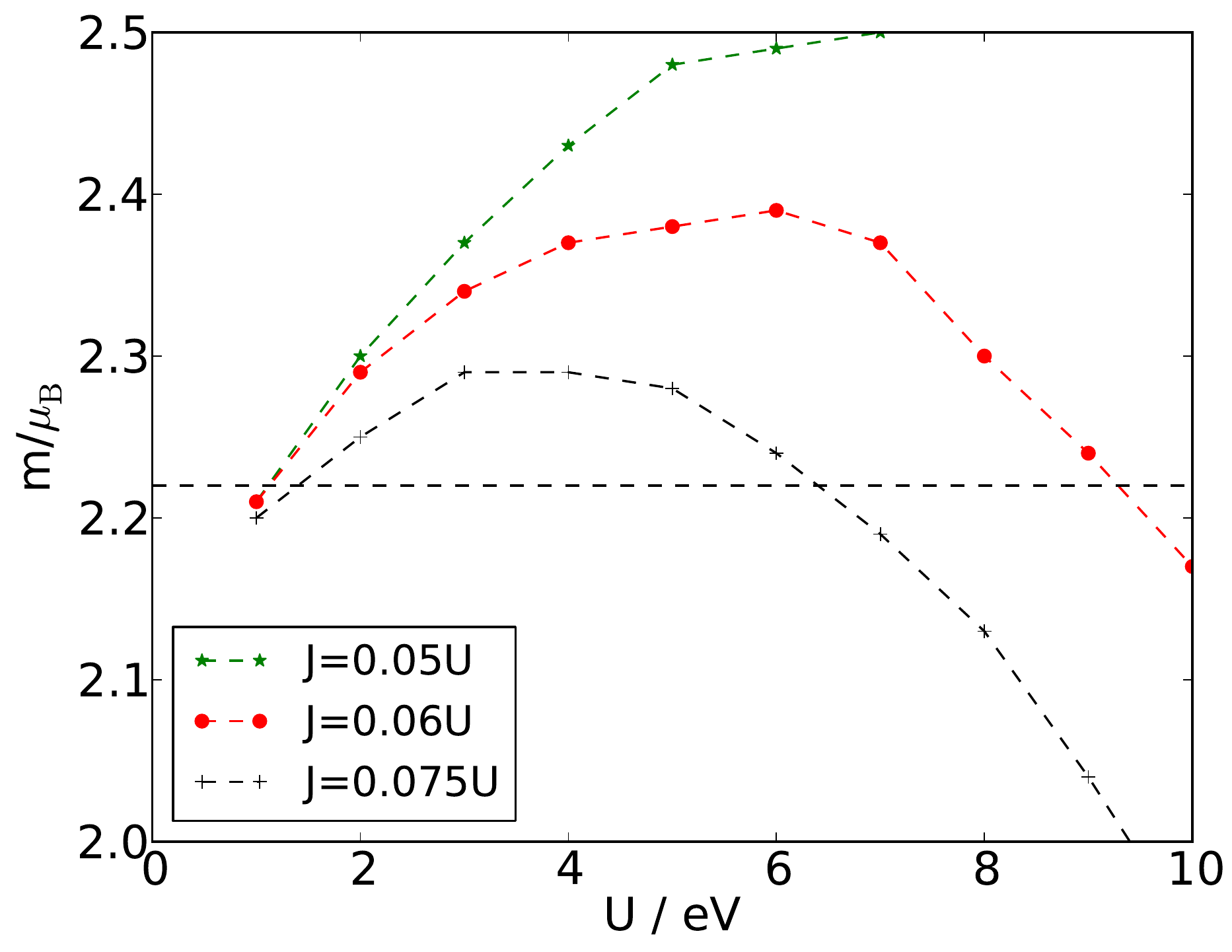}
\caption{Magnetization~$m(U,J)$ in units of Bohr magneton~$\mu_{\rm B}$
as a function of the Hubbard interaction~$U$ for $J/U=0.05,0.06,0.075$.
The horizontal dashed line indicates the experimental value $m=2.22\mu_{\rm B}$.
\label{fig:magnet}}
\end{figure}

\subsubsection{Magnetization}

In Fig.~\ref{fig:magnet} we show 
the ordered magnetic moment $m(U,J)$ as a function of the Hubbard interaction~$U$
for the previously used ratios $J/U=0.05,0.06,0.075$. 
The horizontal dashed line indicates the experimental value,
$m^{\rm exp}=2.22\mu_ {\rm B}$.~\cite{Dananetal}
The ordered magnetic moment $m$ is calculated from the particle
densities $n_{\sigma}(\vecr)$ as
\begin{equation}
m/\mu_{\rm B}= \int d\vecr [n_{\uparrow}(\vecr)-n_{\downarrow}(\vecr)] \,,
\end{equation}
where we used $g=2$ as the electrons' gyromagnetic factor.

It is important to note that the DFT(LDA) 
predicts a large magnetization, i.e., iron is a band ferromagnet in DFT(LSDA).
Figure~\ref{fig:magnet} shows that 
the Coulomb corrections due to the intra-atomic correlations
in the $3d$-shell amount to only 10\% of the magnetization. Indeed, 
in the parameter regime shown in Fig.~\ref{fig:magnet}, 
we have $2.0\mu_{\rm B}< m<2.4\mu_{\rm B}$
for $1\, {\rm eV} \leq U \leq 9\, {\rm eV}$ and $J/U=0.06,0.075$.

Since the Coulomb correlations in the $3d$-shell 
are not the primary cause for magnetism, 
the magnetization $m(U,J)$ does not show a
simple dependence on the Hubbard interaction~$U$ in combination with
the Hund's-rule coupling~$J$. In iron, for vanishing Hund's-rule coupling, $J/U=0$,
we find that the magnetization increases as a function of~$U$,
as also seen in LDA+$U$. This is the usual Stoner mechanism:
in a magnetized system there is less Coulomb energy to be paid, 
at the price of a loss in kinetic energy. When we increase~$U$,
the energy balance is shifted towards the exchange-energy gain so that the
magnetization increases. 

Within the Gutzwiller-DFT, the Hund's-rule coupling leads to
the rather unexpected behavior seen in Fig.~\ref{fig:magnet}.
For fixed~$U$, an {\sl increase\/} of the Hund's-rule coupling~$J$ leads to
a {\sl decrease\/} of the magnetization~$m(U,J)$.
Such a behavior was observed previously in iron,~\cite{Deng2009}
and also in nickel.~\cite{1367-2630-16-9-093034} Moreover,
the influence of the Hund's-rule coupling is not small. 
Indeed, as seen in Fig.~\ref{fig:magnet}, 
it leads to a parabolic downturn of $m(U,J)$ as a function of~$U$
for fixed~$J/U$. We shall discuss the effect of the Hund's-rule
coupling in more detail in Sect.~\ref{sec:moments}.

Using the information in figures~\ref{fig:lat-const} and~\ref{fig:magnet} 
we can determine the optimal values for the interaction parameters. For 
$U_{\rm opt}=9.0\,{\rm eV}$ and $J_{\rm opt}=0.54\,{\rm eV}$, we obtain good results
for the lattice parameter and the magnetic moment,
$a_{\rm opt}=a(U_{\rm opt},J_{\rm opt})=5.39a_{\rm B}=2.85\, \hbox{\AA}$ and 
$m_{\rm opt}=m(U_{\rm opt},J_{\rm opt})=2.24\mu_{\rm B}$, that agree very well 
with the experimental values. 
In the rest of paper we refer to the parameter set $(U_{\rm opt},J_{\rm opt})$
as our `optimal' atomic parameters.

\subsubsection{Size of optimal atomic parameters}
\label{lilia}

Before we proceed, we briefly comment on 
our optimal Coulomb parameters because they
are substantially larger than parameters used in
other studies for 
iron.~\cite{0953-8984-26-37-375601,PhysRevB.81.045117,PhysRevLett.106.106405,%
Leonov2014,PhysRevLett.87.067205}
In most previous studies, 
the values $U = 2\,{\rm eV}\ldots 3\,{\rm eV}$ 
and $J=0.8\,{\rm eV} \ldots 1.0\,{\rm eV}$ are used, e.g., to describe
the high-temperature regime with the transition
from fcc iron to bcc iron and the Curie transition
from non-magnetic to magnetic bcc iron,
while more recent LDA+DMFT studies 
employ larger values, 
$\bar{U}=4.3\, {\rm eV}$, $\widetilde{J}=1.0\,{\rm eV}$.~\cite{PhysRevB.90.155120} 
In all cases, the explored parameter regime appears to be 
quite different from ours.

First of all, we note that the large spread of values of $(U,J)$ 
in the literature is due to the strong sensitivity
of these parameters to the energy window used for
projecting, or {\sl downfolding}, the full electronic structure
to an effective many-body model.~\cite{PhysRevB.80.155134}
It is well known that the bare Hubbard parameters~$U$ are of the order
of~20 eV, or larger.~\cite{Hubbard1963} 
They apply for instantaneous charge excitations
of an isolated atom, which are strongly screened in a solid. In Fe,
for example, the screening reduces $U$ to $\sim$ 3 eV 
for $d$-only models.~\cite{Schickling2011,Schickling2012}
Our self-consistent DFT method is based on a projective technique to
construct Wannier functions. In the present calculations, we chose a large
energy window, which ensures a very good localization of the Fe $3d$
orbitals, and a minimal dependence of the basis set on atomic positions. 
This large energy window translates 
into larger values of $U,J$.~\cite{PhysRevB.74.125106} 
Other calculations can typically afford to retain fewer bands.

Second, we note that the Hubbard-$U$ in our treatment
parameterizes the interaction of two electrons in the same orbital,
see appendix~\ref{app:Coulomb}.
In other approaches, this quantity describes some orbital average.
For example, Pourovskii et al.~\cite{PhysRevB.90.155120} use the Slater-Condon 
parameter~$F^{(0)}=\bar{U}$,
where $\bar{U}=(U+4U')/5$, see eq.~(\ref{eqApp:barU}), and $U'=A-B+C=U-2J$ 
is the inter-orbital Coulomb repulsion.
Naturally, the intra-orbital~$U$ is larger than an average
over intra-orbital and inter-orbital Coulomb repulsions.
Likewise, we work with the average Hund's-rule coupling $J=5B/2+C$,
see eq.~(\ref{eq:appJdef}), 
whereas $\widetilde{J}\equiv(F^{(2)}+F^{(4)})/14=
7B/2+7C/5=7J/5$.~\cite{PhysRevB.90.155120}
Therefore, $F^{(0)}=4.3\, {\rm eV}$ and $\widetilde{J}=1.0\, {\rm eV}$
correspond to $J=0.71\, {\rm eV}$ and 
$U=\bar{U}+8\widetilde{J}/7= 5.4\, {\rm eV}$ with $J/U=0.13$.
We note in passing that 
we work with $C/B=4$ whereas
others use $F^{(2)}/F^{(4)}=8/5$
which corresponds to $C/B=175/47\approx 3.7$.~\cite{PhysRevB.42.5459}

Lastly, in our Gutzwiller calculations, 
we use parameters such as $U$ and~$J$
to `match' selected experimental quantities.
In this way, we compensate approximations in the model setup, e.g., 
the neglect of non-local correlations in Hubbard-type models, 
and in the model analysis, e.g., the limit of infinite dimensions or
an approximate variational ground state. 
For example, in Gutzwiller calculations,
the optimal Coulomb parameters must be chosen somewhat smaller 
when the full atomic interaction 
is replaced by density-density interactions only.~\cite{Schickling2012}
Similarly, larger $U$-values are found to be optimal when the impurity solver
in Quantum-Monte-Carlo is rotationally invariant.~\cite{0953-8984-26-37-375601}
In the following we will show that our optimal
atomic parameters lead to a good agreement with experiment. 
In particular, our substantial Hubbard interaction leads to noticeable
bandwidth renormalizations and an increase of the quasi-particle masses
at the Fermi energy, 
as seen in  experiment.~\cite{PhysRevB.72.155115,PhysRevLett.103.267203}

We note that the atomic parameters for our study of iron
resemble those used in recent LDA+Gutzwiller studies 
by Deng et alii,~\cite{Deng2008, Deng2009}
and our results agree quite well;
on the other hand,
we do not agree with Borghi et al.~\cite{PhysRevB.90.125102}
who advocate small Hubbard interactions in their LDA+Gutzwiller work;
however, as discussed in the following,
there are sizable discrepancy between their and our results 
already at the DFT(LDA) level, and this prevents a detailed comparison.

\subsection{Physical properties within Gutzwiller-DFT}

After fixing the parameters, we are in the position to test the Gutzwiller-DFT against
independent experimental observations. Here, we choose
the bulk modulus and the transition from ferromagnetic bcc iron
to non-magnetic hcp iron. Furthermore, we discuss the local occupancies
in more detail to elucidate the unexpected effect of the
Hund's-rule coupling on the magnetization seen in Fig.~\ref{fig:magnet}.

\begin{figure}[ht]
\includegraphics[width=0.42\textwidth]{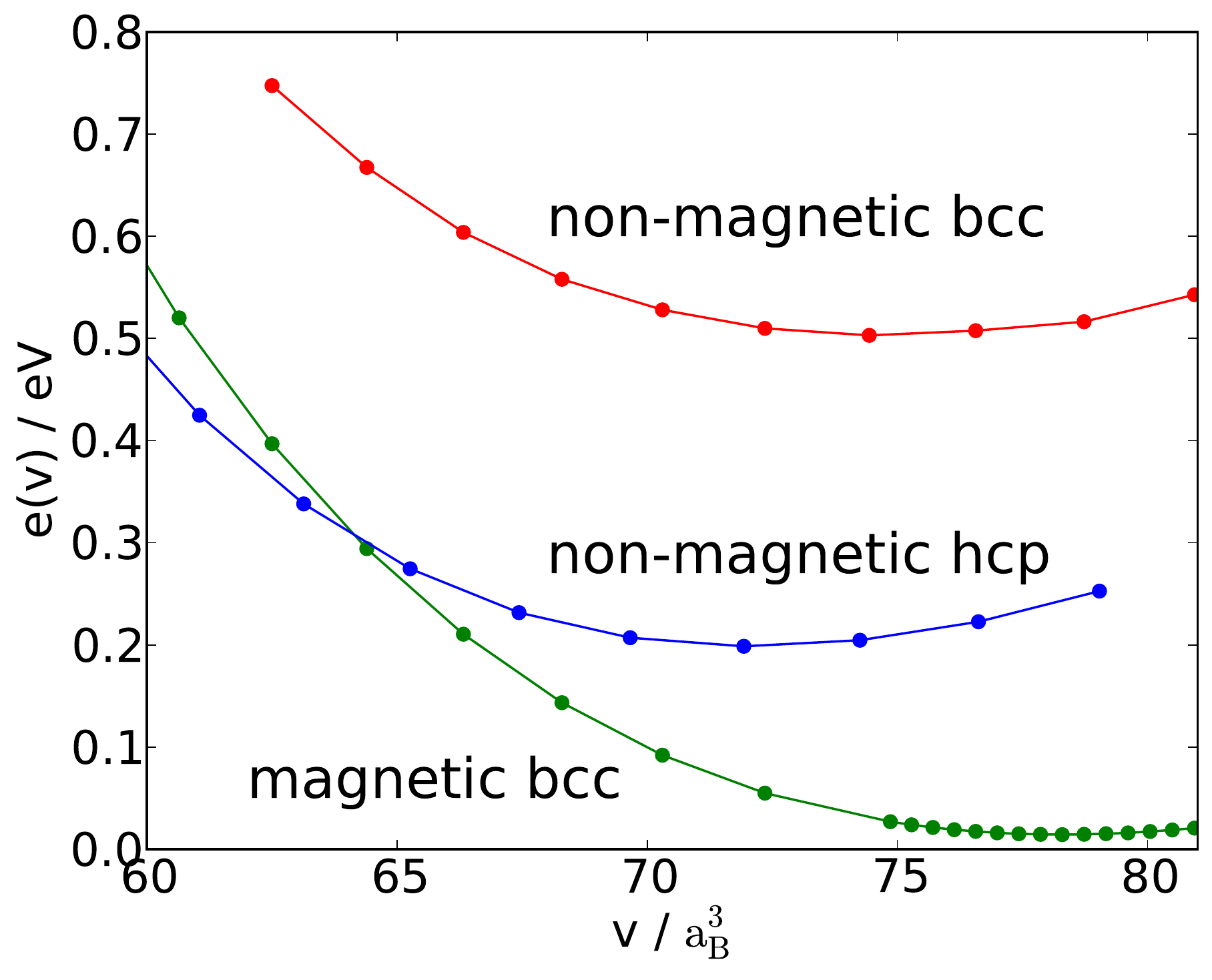}
\caption{Energy per atom $e(v)$ in units of eV as a function of the unit-cell volume~$v$
in units of $a_{\rm B}^3$
for non-magnetic and ferromagnetic bcc iron and non-magnetic hcp 
iron at $U=9\, {\rm eV}$ and $J=0.54\, {\rm eV}$ and ambient pressure.
The energies are shifted by the same constant amount.\label{fig:bulkmodulus}}
\end{figure}

\subsubsection{Bulk modulus}

In Fig.~\ref{fig:bulkmodulus} we show the ground-state energy 
per atom, $e(v)=E(V)/N$, 
as a function of the unit-cell volume~$v=V/N=a^3/2$ in the vicinity of the optimal
value $v_0=a_{\rm opt}^3/2=78.3a_{\rm B}^3=11.6\, \hbox{\AA}^3$ 
with $a_{\rm opt}=5.39a_{\rm B}=2.85\, \hbox{\AA}$.
The bulk modulus at zero temperature 
is defined as the second-derivative of the ground-state energy~$E(V)$
with respect to the volume,
\begin{equation}
B=v_0 \left. \frac{{\rm d}^2 e(v)}{ {\rm d} v^2 }\right|_{v=v_0} \; .
\end{equation}
This implies the Taylor expansion
$e(v)=e(v_0)+(B v_0/2) (v/v_0-1)^2+\ldots $ for the ground-state
energy (Birch-Murnaghan fit). Therefore, we find the bulk modulus from the curvature
of $e(v)$ near $v=v_0$.

In Gutzwiller-DFT we find a bulk modulus of $B=165\,{\rm GPa}$,
in very good agreement with the experimental value,
$B^{\rm exp}=(170\pm 4)\,{\rm GPa}$.~\cite{JGRB:JGRB5666,PhysRevB.82.132409}
The LDA+Gutzwiller value 
substantially improves the DFT(LDA) value of $B^{\rm LDA}=227\,{\rm GPa}$,
it is slightly better than the values from DFT(GGA) studies, 
$B^{\rm GGA}=(190\pm 10)\,{\rm GPa}$,~\cite{PhysRevB.82.132409}
and agrees with the value obtained in DMFT calculations,
$B^{\rm DMFT}=168\,{\rm GPa}$.~\cite{PhysRevB.90.155120}

\subsubsection{Pressure-induced transition from bcc to hcp iron}

Figure~\ref{fig:bulkmodulus} shows that the bcc structure is only stable
because it is ferromagnetic.~\cite{Andersenphysica}
By reducing the volume by applying external pressure,
a first-order structural transition is observed at
a pressure of $p_{\rm c}^{\rm exp}=10\ldots 15\,{\rm GPa}$
at room temperature,~\cite{bcc-to-hcp-first}
together with the concomitant electronic and magnetic 
changes.~\cite{Iota,PhysRevLett.110.117206}

\begin{figure}[ht]
\includegraphics[width=0.44\textwidth]{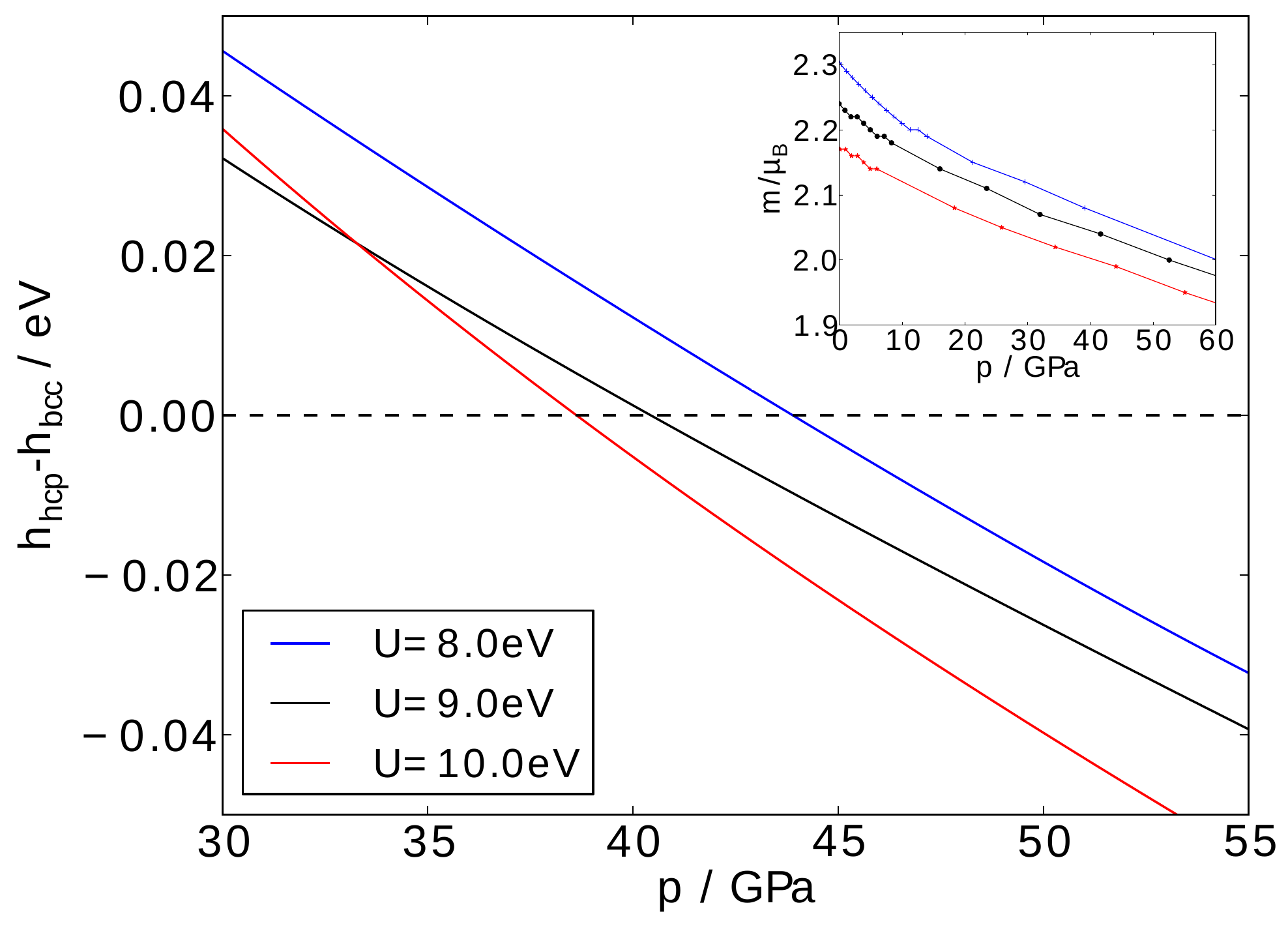}
\caption{Enthalpy difference per atom $h_{\rm hcp}-h_{\rm bcc}$ 
in units of eV between non-magnetic hcp iron
and ferromagnetic bcc iron as a function of applied pressure~$p$
for $U=8.0,9.0,10\, {\rm eV}$ and $J/U=0.06$.
Inset: magnetization as a function of pressure.
\label{fig:enthalpy}}
\end{figure}

In Fig.~\ref{fig:enthalpy} we plot
the enthalpy difference per atom between the non-magnetic hcp lattice
and the ferromagnetic bcc lattice as a function of applied pressure
for $U=8.0,9.0,10\, {\rm eV}$ for fixed ratio $J/U=0.06$.
For our optimal parameter set ($U=9.0\,{\rm eV}, J=0.06U=0.54\, {\rm eV})$ 
we obtain $p_{\rm c}=41\,{\rm GPa}$ as critical pressure at zero temperature,
in qualitative agreement with experiment.
The critical parameter only slightly depends on the value of~$U$
in the vicinity of~$U=U_{\rm opt}$.
We find at $J/U=0.06$ that $p_{\rm c}$ decreases as a function of~$U$,
from $p_{\rm c}=44\,{\rm GPa}$ for $U=8.0\, {\rm eV}$
down to $p_{\rm c}=38\,{\rm GPa}$ for $U=10\, {\rm eV}$.
Therefore, the transition at positive pressures is a robust feature
in Gutzwiller-DFT. We note in passing that the critical pressure sensitively depends
on the ratio $J/U$. For $U=9\, {\rm eV}$, we find $p_{\rm c}=18\, {\rm GPa}$
for $J/U=0.075$. In this case, the magnetization at ambient pressure 
is smaller than in experiment, $m=2.05\mu_{\rm B}$, and, correspondingly, it
requires less pressure to destroy the ferromagnetic bcc ground state.

Our Gutzwiller-DFT values for $p_c$ are larger than the experimental values
observed at room temperature.  
Our calculation applies to zero temperature while, at finite temperatures,
phonon, magnon, and electronic quasi-par\-ticle contributions
to the entropy also add to the difference
in the Gibbs' free energies between the two phases.
The latter contributions may not be unimportant because 
the magnetic order is destroyed at the transition.
Whether the bcc or the hcp phase is stabilized by the various
entropy contributions is unresolved.

We note in passing that, for simplicity, we have done the hcp calculations with 
the same local Hamiltonian as used for the bcc calculations, 
see appendix~\ref{app:Coulomb}.
The latter explicitly uses cubic symmetry. Since we work in spherical
approximation in any case, we do not expect that this additional
approximation induces significant corrections.

In the inset of Fig.~\ref{fig:enthalpy} we show the  
magnetization in bcc iron as a function of pressure when we ignore the
structural transition. The magnetization changes by less than 10\%
from ambient pressure to $p_{\rm c}$, and it would vanish at much large pressures,
$p_{\rm m}>600\, {\rm GPa}\gg p_{\rm c}$. 
Therefore, we find that the first-order transition at $p_{\rm c}$ is not triggered
by a collapse of the magnetization in bcc iron.

\subsubsection{Local occupancies}
\label{sec:moments}

In iron, the atomic $4s$~electrons strongly hybridize 
with the $3d$~levels and increase their average
occupancy from the atomic value $\overline{n}_d^{\rm atom}=6$ 
to $\overline{n}_d^{\rm LDA}=7.3$ in DFT(LDA).
The double-counting correction
used in this work, see appendix~\ref{app:Coulomb},
keeps~$\overline{n}_d$ essentially constant. 
We find  $\overline{n}_d^{\rm GDFT}=7.2$ for $U=9\, {\rm eV}$
and $J/U=0.06$. Therefore, the local Coulomb interactions
merely redistribute the electrons among the 1024 atomic $3d$~configurations.

\begin{figure}[b]
\includegraphics[width=0.40\textwidth]{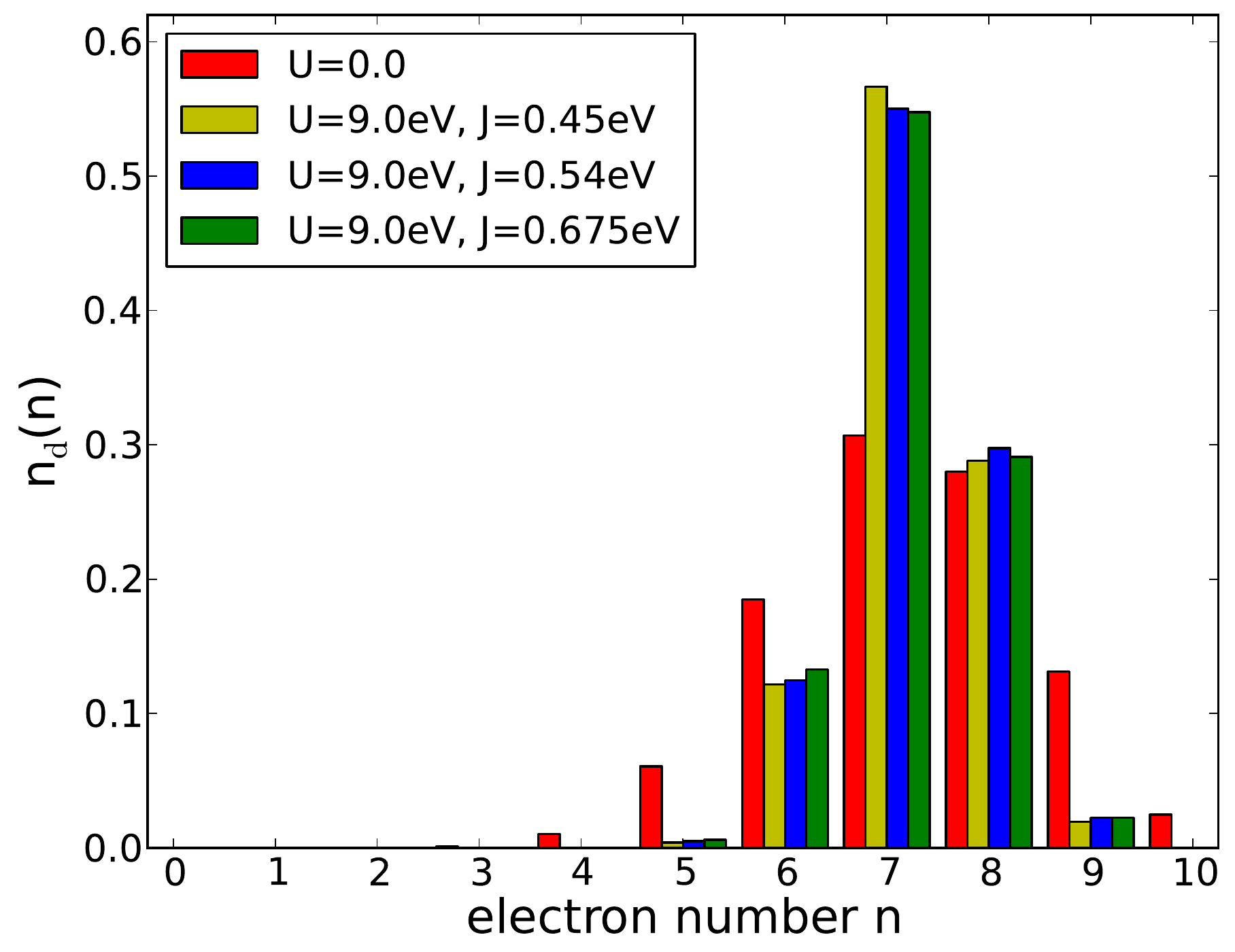}
\caption{Local charge distribution of iron atoms $n_d(n)$
for $U=0$ (LDA limit)
and $U=9\, {\rm eV}$ and $J/U=0.05,0.06,0.075$ (Gutzwiller-DFT)
for optimal lattice parameters.\label{fig:locnop}}
\end{figure}

The average $3d$-electron density~$n_d$ and the 
magnetization~$m$ do not change much as a function of $(U,J)$.
However, this does not imply that correlations are small in iron.
In order to display the correlated nature of the ground state,
we study some local properties.

First, we discuss the local charge distribution~$n_d(n)$
which gives the probability to find $n$~electrons in the $3d$~shell on an iron atom.
Figure~\ref{fig:locnop} shows $n_d(n)$ from Gutzwiller-DFT in the LDA limit,
$U=0$, and for $U=9\, {\rm eV}$ and $J/U=0.05,0.06,0.075$.
For $U=0$ we find quite a broad distribution $n_d(n)$ with 
significant values for $n_d(n)$ for $4\leq n\leq 10$.
For $U=9.0\,{\rm eV}$ and 
$J/U=0.05,0.06,0.075$, only configurations with $n=6,7,8$ electrons
in the $3d$~shell have a substantial weight. This does not come as a surprise because
the Gutzwiller correlator suppresses the occupation
of local configurations that are energetically unfavorable.
This behavior was also
observed in previous studies.~\cite{Schickling2011,Schickling2012}

\begin{figure}[htb]
\includegraphics[width=0.40\textwidth]{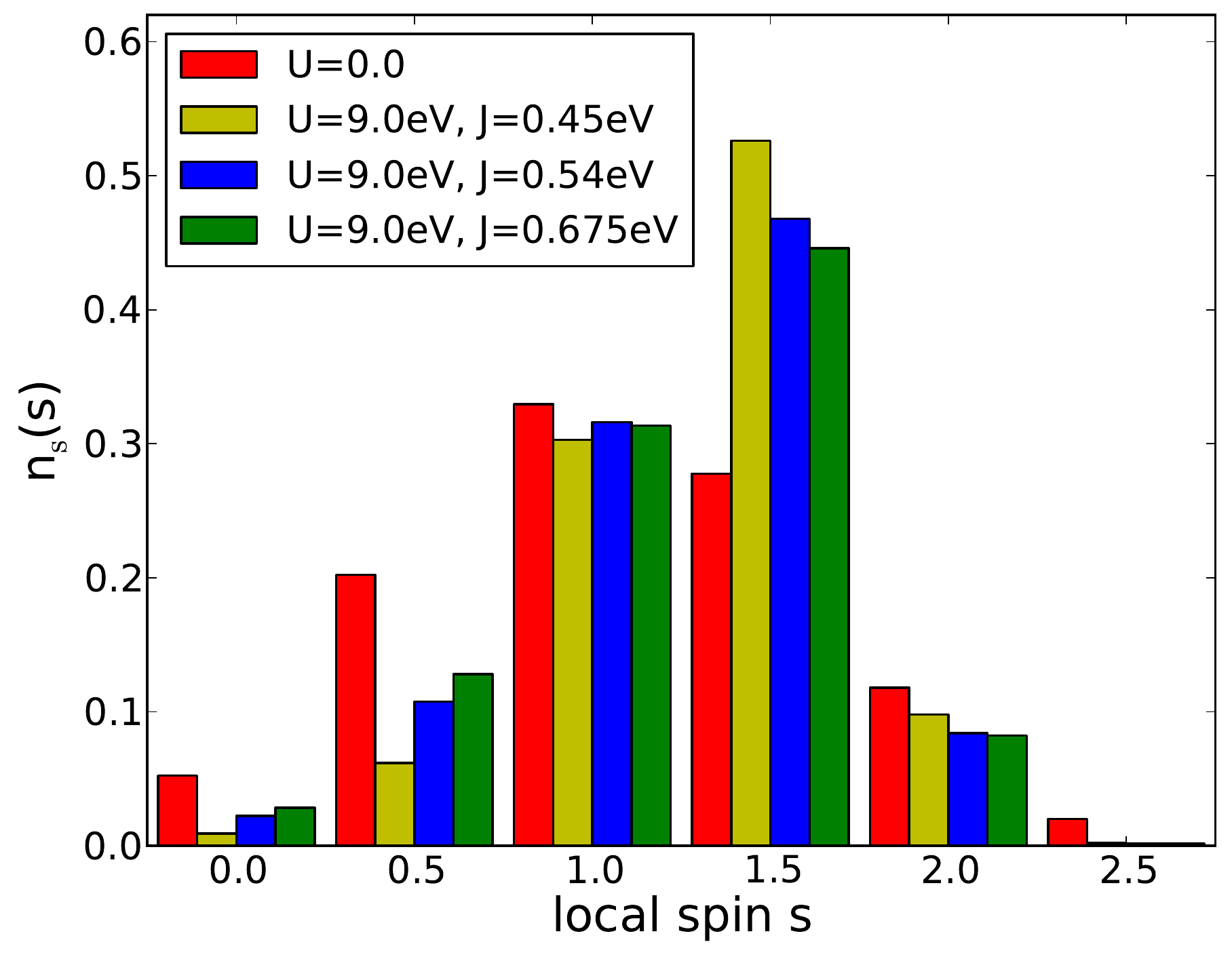}
\caption{Local spin distribution of iron atoms $n_s(s)$ 
for \hbox{$U=0$} (LDA limit)
and $U=9\, {\rm eV}$ and $J/U=0.05,0.06,0.075$ (Gutz\-willer-DFT)
for optimal lattice parameters.\label{fig:locspin}}
\end{figure}

More revealing is the local spin distribution function~$n_s(s)$  shown
in Fig.~\ref{fig:locspin} where $n_s(s)$ gives the probability
to find the local spin quantum number~$s$ on an iron atom. From $n_s(s)$
we can calculate the expectation value for the local spin as
\begin{equation}
\langle \hat{\vecS}_i^2 \rangle = \sum_s n_s(s) s(s+1) \equiv S_m(S_m+1)\;.
\label{eqn:S}
\end{equation}
Here, $S_m$ defines our average spin per atom.
Figure~\ref{fig:locspin} shows that the local spin distribution 
is fairly broad for $U=0$. Since $n_d=5$ is finite for $U=0$, the local
spin distribution has a finite weight even for $s=5/2$.
For $U=9\, {\rm eV}$, the local configurations $3d^7$ and $3d^8$ 
dominate, see Fig.~\ref{fig:locnop}.
Therefore, applying Hund's first rule,
we expect to find peaks in the local spin distribution $n_s(s)$
at $s=3/2$ and $s=1$ which is indeed seen in Fig.~\ref{fig:locspin}.
Concomitantly, the average spin per atom is a bit larger for $U=9\, {\rm eV}$,
$S_m= 1.3$ for $J/U=0.06$, than for $U=0$, $S_m^{\rm LDA}=1.23$.

In contrast to Hund's first rule, $S_m(J)$ {\sl decreases\/} as a function
of the Hund's-rule coupling~$J$. This is seen from Fig.~\ref{fig:locspin}
which shows that the weight of configurations with spin $s=0,1/2$ increases
at the expense of configurations with spin $s=3/2$.
Therefore, both the overall magnetization~$m$ and the local spin~$S_m$
decrease as a function of the Hund's-rule coupling.
This seems to contradict Hund's first rule which states that, in an atom, a larger~$J$ 
stabilizes configurations with a larger spin. Apparently, the solution to this
problem must be related to the fact that we are investigating a
metal in which band-magnetism dominates.

In table~\ref{table:en-and-loc-quan}
we list the values for several
quantities for $J/U=0.05$, $J/U=0.06$ and $J/U=0.075$ at
fixed $U=9.0\,{\rm eV}$ and fixed lattice parameter $a=5.39a_{\rm B}$.
The data redisplay 
the behavior seen in figures~\ref{fig:magnet} and~\ref{fig:locspin}:
when we increase $J/U$, the magnetization~$m$ decreases.
Note, however, that upon an increase of~$J/U$, 
the electronic correlations actually increase, too,
as can be seen from the bandwidth reduction factors~$(q^2)$.
The itineracy of the electrons becomes progressively
worse when the weight of local configurations
is redistributed by the Gutzwiller correlator for increasing~$J/U$.
The effect of the Hund's-rule coupling on the bandwidth reduction
is fairly pronounced. The $(q^2)$-factors  decrease by $\Delta (q^2) \approx 0.2$ when
we vary~$U$ from zero to $U=9.0\,{\rm eV}$ but they
change by as much as $\Delta (q^2) \approx 0.1$ for the majority spin species
when we go from $J/U=0.05$ to $J/U=0.075$ at $U=9\, {\rm eV}$.
Apparently, it is favorable for the kinetic energy to flip majority spins back to
minority spins, i.e., there is a tendency to {\sl reduce\/} the magnetization
as a function of~$J/U$.
As seen from table~\ref{table:en-and-loc-quan}, 
the partial occupancy $n_{e_g}$ of the $e_g$-levels remains almost unchanged, and
the reduction of the magnetization from $m=2.49\mu_{\rm B}$ at $J/U=0.05$ to
$m=2.05\mu_{\rm B}$ 
at $J/U=0.075$ is generated by flipping majority-spin $t_{2g}$-electrons.

\begin{table}[t]
\caption{Magnetization~$m$, 
bandwidth reduction factors $(q_{e_g})^2$ and $(q_{t_{2g}})^2$ 
for $3d(e_g)$ and $3d(t_{2g})$ electrons, respectively
(resolved for majority spins and minority spins),
kinetic energy $E_{\rm kin}$ per atom, and interaction energy $E_{\rm int}$
per atom; Gutzwiller-DFT data for $U=9.0\,{\rm eV}$ for $J/U=0.05,0.06,0.075$.
The lattice parameter is fixed at $a=5.39 a_{\rm B}$.\label{table:en-and-loc-quan}}
\begin{tabular*}{0.48\textwidth}{l @{\extracolsep{\fill}}ccc}
		\hline
		\hline
		& \vphantom{\Large A}$J/U=0.05$ & $J/U=0.06$ & $J/U=0.075$ \\
		\hline
		$m/\mu_{\rm B}$ & 2.49 & 2.24 & 2.05\vphantom{\Large A} \\
		$(q_{e_g})^2$ & 0.794, 0.799 & 0.748, 0.799 & 0.712, 0.795 \\
		$(q_{t_{2g}})^2$ & 0.790, 0.783 & 0.746, 0.779 & 0.718, 0.770 \\
		$n_{e_g}$ & 0.985, 0.350 & 0.985, 0.341 & 0.985, 0.332 \\
	 	$n_{t_{2g}}$ & 0.970, 0.546 & 0.925, 0.594 & 0.887, 0.632 \\
		$E_{\rm kin}$/eV & 94.82 & 94.90 & 95.05 \\
		$E_{\rm int}$/eV & 1.68 & 1.48 & 1.16 \\[2pt]
		\hline
		\hline
\end{tabular*}
\end{table}

The Hund's-rule coupling~$J$ changes the weight of iso-electronic local configurations
with different spin. Apparently, this level splitting impedes the average electron transfer
between atoms much more than the elimination of charge states with $n=4,5,9,10$
by the Hubbard interaction~$U$. The loss in kinetic energy
by this  `configurational hopping blockade'
cannot be compensated fully by a gain in local interaction energy
that is at most of the order of~$\Delta E_J=
(J_1S_{m,1}-J_2S_{m,2})$ with $\Delta E_J \approx 0.2\,{\rm eV}$
per atom in our example. Instead, the system prefers to re-gain kinetic energy
by reducing the magnetization at the price of loosing exchange energy; 
recall that in a band magnet 
the loss in kinetic energy is compensated by the gain in exchange energy.
Since all quantities are determined self-consistently in Gutzwiller-DFT, 
the kinetic energy, the exchange energy, and the gain in Hund's-rule energy
must be newly balanced to re-adjust~$m$ 
when we change~$J/U$ for fixed~$U$. 
Apparently, in band magnets we observe a intricate
interplay between atomic and bandstructure physics.

\section{Bandstructure}
\label{sec:bands}

\subsection{Bandwidth renormalization}

We begin with a comparison of the quasi-particle bands from
DFT(LDA), DFT(GGA) and Gutzwiller-DFT with $U=9\, {\rm eV}$ and $J/U=0.06$
for ferromagnetic bcc iron 
at ambient pressure ($a=5.39a_{\rm B}$).
Moreover, we compare bands from DFT(LDA) and Gutzwiller-DFT for
non-magnetic hcp iron with lattice parameter $a_{\rm hcp}=4.60a_{\rm B}$
and $(c/a)_{\rm hcp}=1.60$; the results change marginally when we use
the ideal ratio $(c/a)_{\rm ideal}=\sqrt{8/3}=1.63$.

In order to obtain smooth band plots and to include the effects of
the spin-orbit coupling on the bandstructure perturbatively, 
see Sect.~\ref{sec:so-inclusion}, we introduced a further
post-processing step. The Gutzwiller Kohn-Sham quasi-particles 
were used to generate maximally-localized
Wannier functions using {\sc Wannier90},~\cite{Mostofi2008685}
from which we constructed a tight-binding model to calculate 
the band structure at arbitrary  ${\bf k}$-points.
These Wannier functions are used only for plotting purposes, and
are unrelated to those chosen to perform the self-consistent calculations.

We checked that the tight-binding dispersion relation agrees with
the calculated energy levels from {\sc Quantum ESPRESSO} 
for our selected independent ${\bf k}$-points in the Brillouin zone.
The small wiggles in the $4s$-bands close to the $\Gamma$ point seen 
in Fig.~\ref{fig:wanbands} are a result of the tight-binding fit.
We disregard the problem because this does not influence  the 
$3d$-bands close to the Fermi energy and has no effect on the total energy 
which is calculated using the original quasi-particles.

\subsubsection{Ferromagnetic bcc iron}

\begin{figure}[t]
\includegraphics[width=0.44\textwidth]{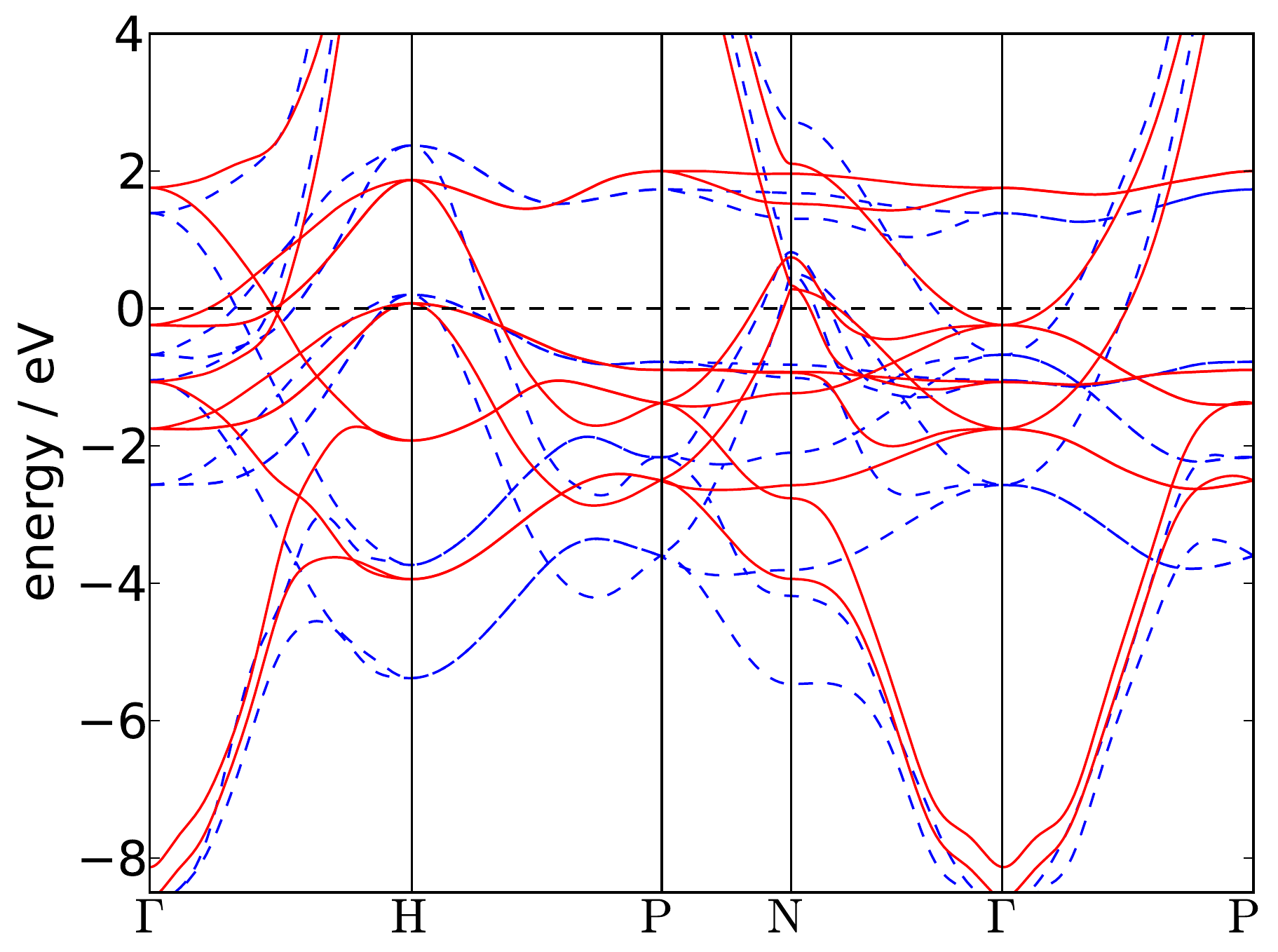}
\caption{Comparison between DFT(LDA) bands (blue, dashed lines) 
for $a^{\rm LDA}=5.21a_{\rm B}$
and bands from Gutzwiller-DFT (red, full lines) 
for the optimal atomic parameters 
$U=9.0\,{\rm eV}$ and $J=0.54\,{\rm eV}$ 
and $a=5.39a_{\rm B}$ for ferromagnetic bcc iron. 
For clarity, we do not discriminate between majority
and minority spin bands. The Fermi energy is at $E_{\rm F}=0$
(dashed black horizontal line).\label{fig:wanbands}}
\end{figure}

In Fig.~\ref{fig:wanbands} we compare the bandstructure for ferromagnetic bcc
iron from DFT(LDA) 
and from Gutzwiller-DFT for $U=9.0\,{\rm eV}$ and $J=0.54\,{\rm eV}$.
Both calculations are performed at the optimal
lattice parameter,  $a^{\rm LDA}=5.21a_{\rm B}$
and $a^{\rm GDFT}=5.39a_{\rm B}$. 
Fig.~\ref{fig:wanbands} shows the common characteristics of correlation-induced 
effects on energy bands. First, the uncorrelated, $4sp$-type parts of the 
quasi-particle bands deep below the Fermi energy
do not differ much, e.g., the lowest $4sp$-type majority bands are 
at $\Gamma_{{\rm low},\uparrow}^{\rm LDA}=9.0\, {\rm eV}$
and $\Gamma_{{\rm low},\uparrow}^{\text{G-DFT}}=8.57\, {\rm eV}$ 
below the Fermi energy $E_{\rm F}=0$.
Minor deviations are related to slightly 
different lattice parameters and $3d$-electron numbers~$\overline{n}_d$
in DFT(LDA) and Gutzwiller-DFT.

\begin{figure}[b]
\begin{center}
\includegraphics[width=0.44\textwidth]{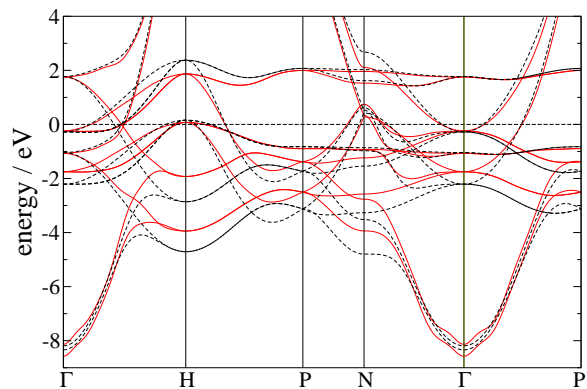}
\end{center}
\caption{Comparison between DFT(GGA) bands (black, dashed lines) 
for $a_{\rm GGA}=5.39a_{\rm B}$
and bands from Gutzwiller-DFT (red, full lines) 
for the optimal atomic parameters 
$U=9.0\,{\rm eV}$ and $J=0.54\,{\rm eV}$ 
and $a=5.39a_{\rm B}$ for ferromagnetic bcc iron. 
For clarity, we do not discriminate between majority
and minority spin bands. The Fermi energy is at $E_{\rm F}=0$
(dashed horizontal line).\label{fig:wanbandsGGA}}
\end{figure}

Second, the Gutzwiller-correlated $3d$-type parts of the quasi-particle bands 
close to the Fermi energy 
are shifted with respect to the DFT(LDA) bands, and the
bandwidths of the correlated bands are reduced by
factors proportional to 
$(q_{t_{2g}})^2$ and $(q_{e_g})^2$
for the $3d$-$t_{2g}$ and $3d$-$e_g$ majority and minority bands.
Note that, due to the hybridization of the quasi-particles, 
a meaningful symmetry
character can only be 
assigned to the bands at high-symmetry points in the Brillouin zone.

The bandwidth reduction in iron is not as strong as in nickel. Nevertheless,
for selected symmetry points, the discrepancies between the 
quasi-particle bands from DFT(LDA) and Gutzwiller-DFT
are quite large.
For example, at the H-point in the Brillouin zone
we find a bandwidth
reduction for the majority band by 36\%, from 
$H_{{\rm low},\uparrow}^{\rm LDA}=5.38\,{\rm eV}$ down to
$H_{{\rm low},\uparrow}^{\text{G-DFT}}=3.94\,{\rm eV}$,
in good agreement with experiment, 
$H_{{\rm low},\uparrow}=3.8\,{\rm eV}$.~\cite{PhysRevB.29.2986}
Likewise, at the N-point in the Brillouin zone
there is a majority spin band at $N_{{\rm low},\uparrow}=4.5\,{\rm eV}$
below the Fermi energy in experiment,~\cite{PhysRevB.29.2986} 
in comparison with 
$N_{{\rm low},\uparrow}^{\rm LDA}=5.47\,{\rm eV}$ in DFT(LDA) and
$N_{{\rm low},\uparrow}^{\text{G-DFT}}=3.90\,{\rm eV}$ in Gutzwiller-DFT.
At the $\Gamma$-point the bandwidth
reduction is only about 10\% for bands close to the Fermi edge.
In addition, the bandwidth renormalization at the $\Gamma$-point
is overlaid with a bandshift of about $0.4\,{\rm eV}$.

For completeness, we show the bandstructure for ferromagnetic
bcc iron from Gutzwiller-DFT for $U=9.0\, {\rm eV}$ and $J=0.54\, {\rm eV}$
in comparison with those from (scalar relativistic) DFT(GGA) calculations
in Fig.~\ref{fig:wanbandsGGA}. The two band structures
differ less than in Fig.~\ref{fig:wanbands} because our DFT(GGA) provides
the same equilibrium lattice parameter as used in Gutzwiller-DFT, 
$a_{\rm GGA}=5.39a_{\rm B}$. 
The bandwidth of the $3d$-electrons from the DFT(LDA)
is calculated for $a_{\rm LDA}=5.21a_{\rm B}$ so that
the $3d$-orbitals have a larger overlap in DFT(LDA) than in DFT(GGA),
and the $3d$ bandwidth is larger in  LDA than in GGA.
Nevertheless, the correlations in the Gutzwiller approach
lead to an additional bandwidth reduction of the $d$ bands across the Brillouin zone.

\begin{figure}[b]
\includegraphics[width=0.44\textwidth]{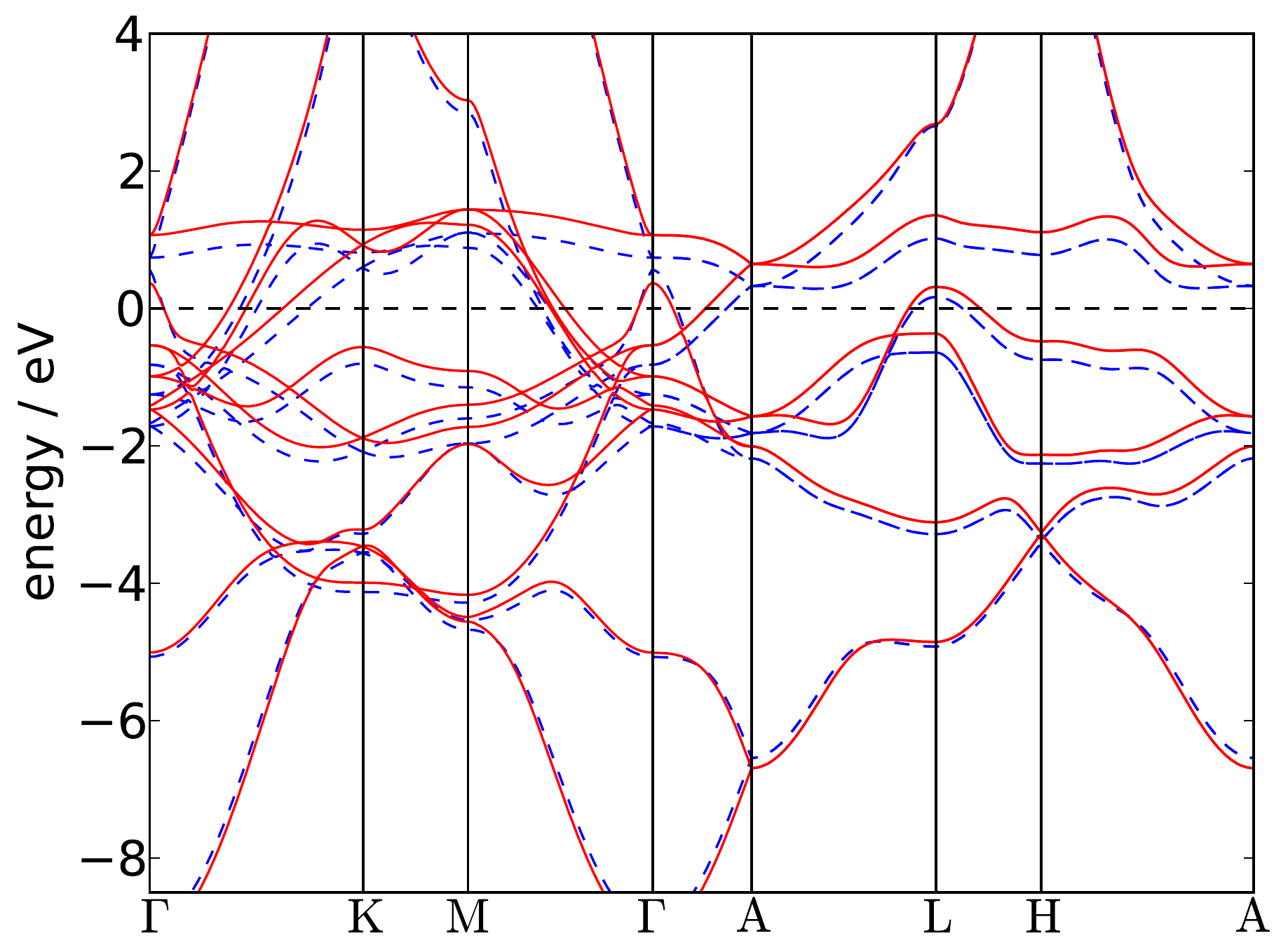}
\caption{Comparison between DFT(LDA) bands (blue, dashed lines) 
and bands from Gutzwiller-DFT (red, full lines) 
for the atomic parameters  $U=9.0\,{\rm eV}$ and $J=0.54\,{\rm eV}$ 
for paramagnetic hcp iron at $a=4.60a_{\rm B}$ and $c/a=1.60$.
The Fermi energy is at $E_{\rm F}=0$
(dashed black horizontal line).\label{fig:wanbands-hcp}}
\end{figure}

\subsubsection{Non-magnetic hcp iron}

In Fig.~\ref{fig:wanbands-hcp} 
we compare the bandstructure for non-magnetic hcp
iron from DFT(LDA) 
and from Gutzwiller-DFT for $U=9.0\,{\rm eV}$ and $J=0.54\,{\rm eV}$.
Both calculations are performed at the lattice parameter $a=4.60a_{\rm B}$
and $c/a=1.60$ so that the unit-cell volume 
is $v_0=\sqrt{3}/2 (a^2/2)c=67.4a_{\rm B}^3$.
The partial densities are almost identical, $\overline{n}_d=7.3$.

As for ferromagnetic bcc iron,  the uncorrelated, $4sp$-type parts of the 
quasi-particle bands deep below or high above the Fermi energy
do not differ much. Again, the Gutzwiller-correlated $3d$-type 
parts of the quasi-particle bands close to the Fermi energy 
are shifted with respect to the DFT(LDA) bands, and the
bandwidths of the correlated bands are reduced. 
The Fermi-liquid properties (Fermi surface topology,  wavevectors, velocities) 
differ only quantitatively.

\renewcommand{\tabularxcolumn}[1]{>{\normalsize\centering\arraybackslash}m{#1}}
\renewcommand{\arraystretch}{1.3}
\begin{table*}
\caption{Comparison of Fermi wavenumbers and velocities 
for various Fermi sheets between 
fully relativistic 
DFT(GGA), 
LDA+Gutzwiller for $U=9.0\,{\rm eV}$, $J=0.54\,{\rm eV}$,
and $a=5.39a_{\rm B}$, and ARPES results.~\protect\cite{PhysRevB.72.155115}  
FS: Fermi sheet; $k_{\rm F}$: Fermi wavenumber; 
$v_{\rm F}=v/\hbar=\hbar k_{\rm F}/m$: Fermi velocity.\label{tble:comp}}
\begin{tabularx}{\textwidth}{XXXXXXXXXXX}
\hline
\hline
Direction & Spin & FS & $k_{\rm F}$ 
\newline GGA \newline$(\hbox{\AA}^{-1})$ & 
$k_{\rm F}$ 
\newline LDA+G \newline$(\hbox{\AA}^{-1})$ &
$k_{\rm F}$ \newline ARPES 
\newline$(\hbox{\AA}^{-1})$ & Slope $v_{\rm D}$ 
\newline GGA
\newline$({\rm eV}\hbox{\AA})$ &
Slope $v_{\rm G}$ 
\newline LDA+G \newline $({\rm eV}\hbox{\AA})$ &
Slope $v_{\rm R}$ \newline ARPES 
\newline $({\rm eV}\hbox{\AA})$  
& Mass \newline Ratio \newline $v_{\rm D}/v_{\rm R}$
& Mass \newline Ratio \newline $v_{\rm G}/v_{\rm R}$
\\
\hline
$\Gamma$--P & Min. & VI & 0.31 & 0.33 & 0.32 & 1.66 &1.38 & 0.88 & 1.9 &1.6\\
           & Maj. & I & 0.95 & 0.94 & 0.97 & 4.74 & 3.86 & 1.40 & 3.4 &2.8 \\
$\Gamma$--H & Min. & VI & 0.47 & 0.49 & 0.46 & 1.08 & 0.83 & 0.72 & 1.5 &1.2 \\
           & Maj. & I & 1.09 & 1.10 & 1.08 & 2.36 & 3.35 & 1.12 & 2.1 & 3.0 \\
               & & II & 1.94 & 1.93  & 1.70  & 0.64 & 0.58 & 0.67 & 1.0 & 0.9 \\
$\Gamma$--N & Min. & VI & 0.33 & 0.35 & 0.36 & 1.52 & 1.25 & 0.80 & 1.9 &1.6 \\
                 & Maj. & I & 1.21 & 1.21 & 1.22 & 1.89 & 1.53  & 1.16 & 1.6 &1.3 \\
H--P & Min. & V & 0.65 & 0.64 & 0.68 & 4.82 & 4.16 & 1.79 & 2.7 &2.3 \\
\hline
\hline
\end{tabularx}
\end{table*}

\subsection{Comparison with ARPES measurements}

\subsubsection{Inclusion of spin-orbit coupling}
\label{sec:so-inclusion}

As effective parameter for the spin-orbit interaction 
we choose $\zeta = 0.06\,{\rm eV}$,
in agreement with previous studies.~\cite{PhysRevB.11.287,PhysRevLett.92.037204}
The small value permits a perturbative treatment
of the spin-orbit coupling. In effect, it leads to negligibly small changes
in the bandstructures but 
for avoided crossings of majority and minority bands where
it induces bandgaps of the order of~$\zeta$.
Since some of the avoided crossings are energetically close
to the Fermi energy, the spin-orbit interaction has some
noticeable effect on the positions of the Fermi points
and the Fermi velocities.

For our perturbative treatment, we start from the majority and minority
bands as calculated from Gutzwiller-DFT for 
($U=9.0\,{\rm eV}$, $J=0.54\,{\rm eV}$) at $a=5.39a_{\rm B}$ and
use the program {\sc Wannier90}~\cite{Mostofi2008685}
to derive a tight-binding Hamilton operator.
Then, the two block-diagonal parts of the
Hamiltonian for majority and minority bands are 
coupled by the spin-orbit interaction.
We obtain the bandstructure with spin-orbit coupling from the
diagonalization of this effective Hamilton matrix.
For larger values of~$\zeta$, a fully self-consistent treatment
of the spin-orbit interaction is necessary that requires
the formulation of a relativistic Gutzwiller-DFT.

\subsubsection{Quasi-particle bands close to the Fermi energy}

\begin{figure}[htb]
\includegraphics[width=0.45\textwidth]{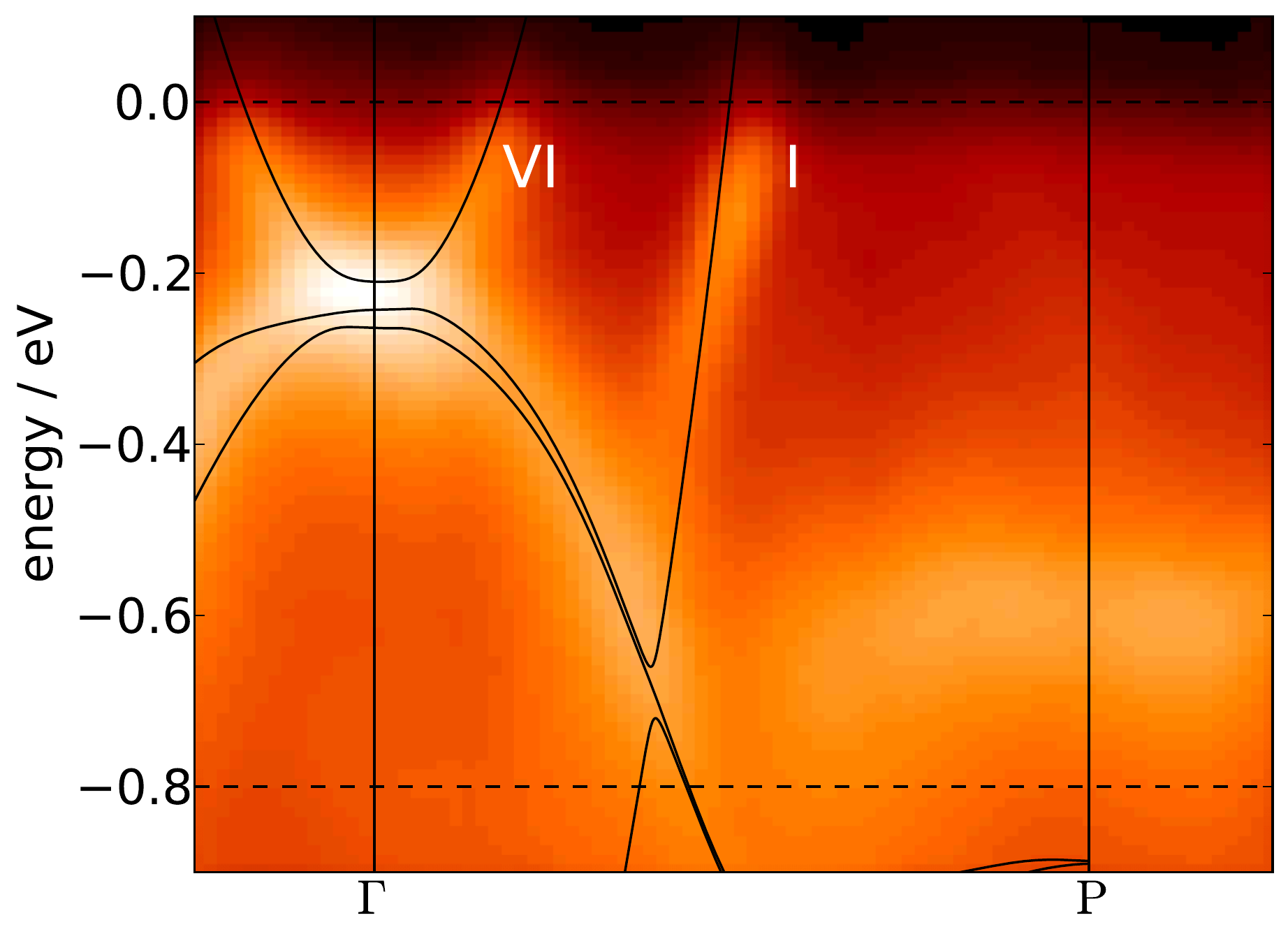}
\caption{Overlay of quasi-particle bands from LDA+Gutzwil\-ler
for $U=9.0\,{\rm eV}$, $J=0.54\,{\rm eV}$, and 
lattice parameter $a=5.39a_{\rm B}$ 
with ARPES data along the $\Gamma$--P 
direction.~\protect\cite{PhysRevB.72.155115}
\label{fig:soc_GP}}
\end{figure}

\begin{figure}[hb]
\includegraphics[width=0.45\textwidth]{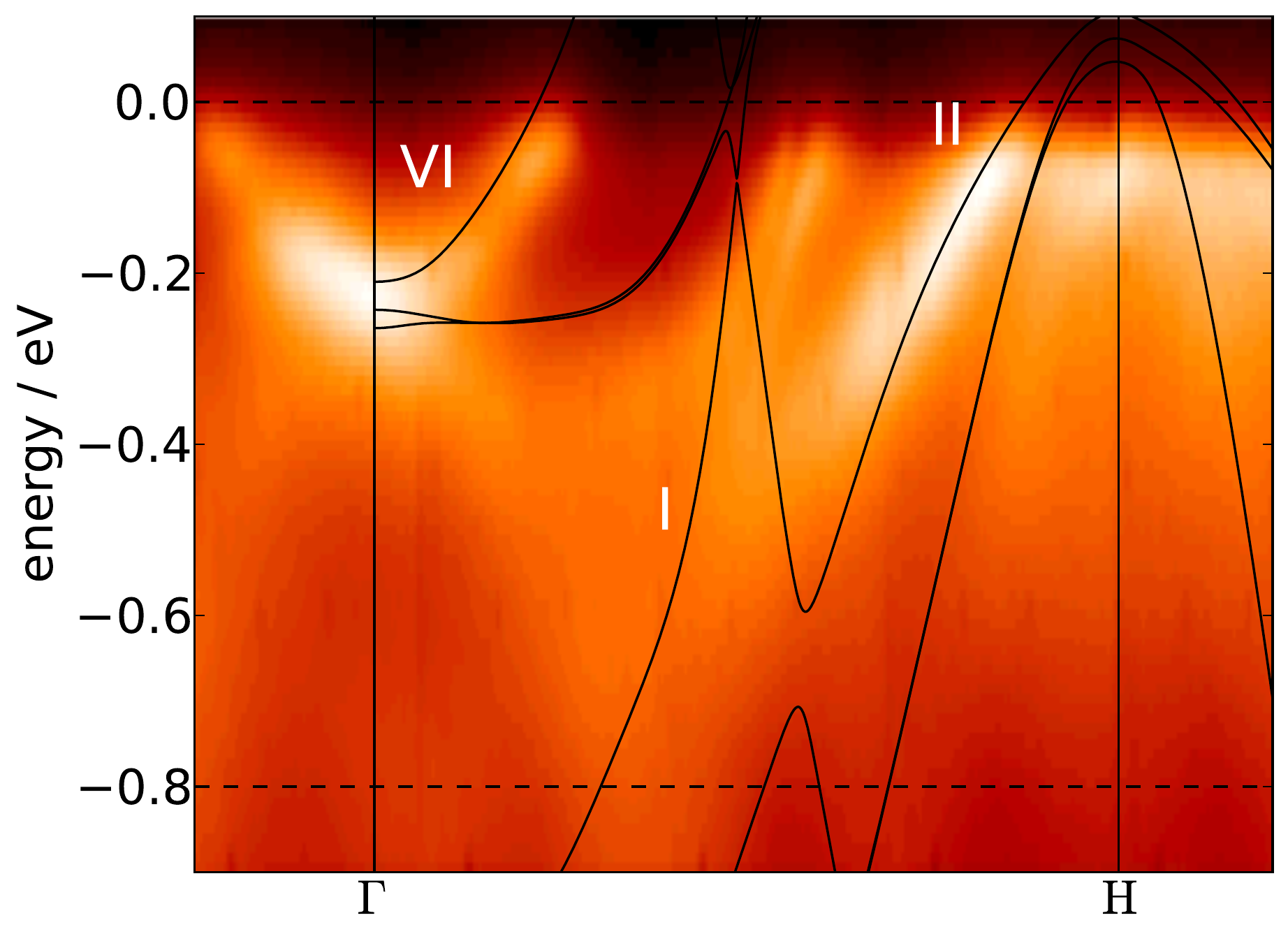}
\caption{Overlay of quasi-particle bands from LDA+Gutzwil\-ler
for $U=9.0\,{\rm eV}$, $J=0.54\,{\rm eV}$, and 
lattice parameter $a=5.39a_{\rm B}$ 
with ARPES data
along the $\Gamma$--H direction.~\protect\cite{PhysRevB.72.155115}
\label{fig:soc_GH}}
\end{figure}

\begin{figure}[hb]
\includegraphics[width=0.45\textwidth]{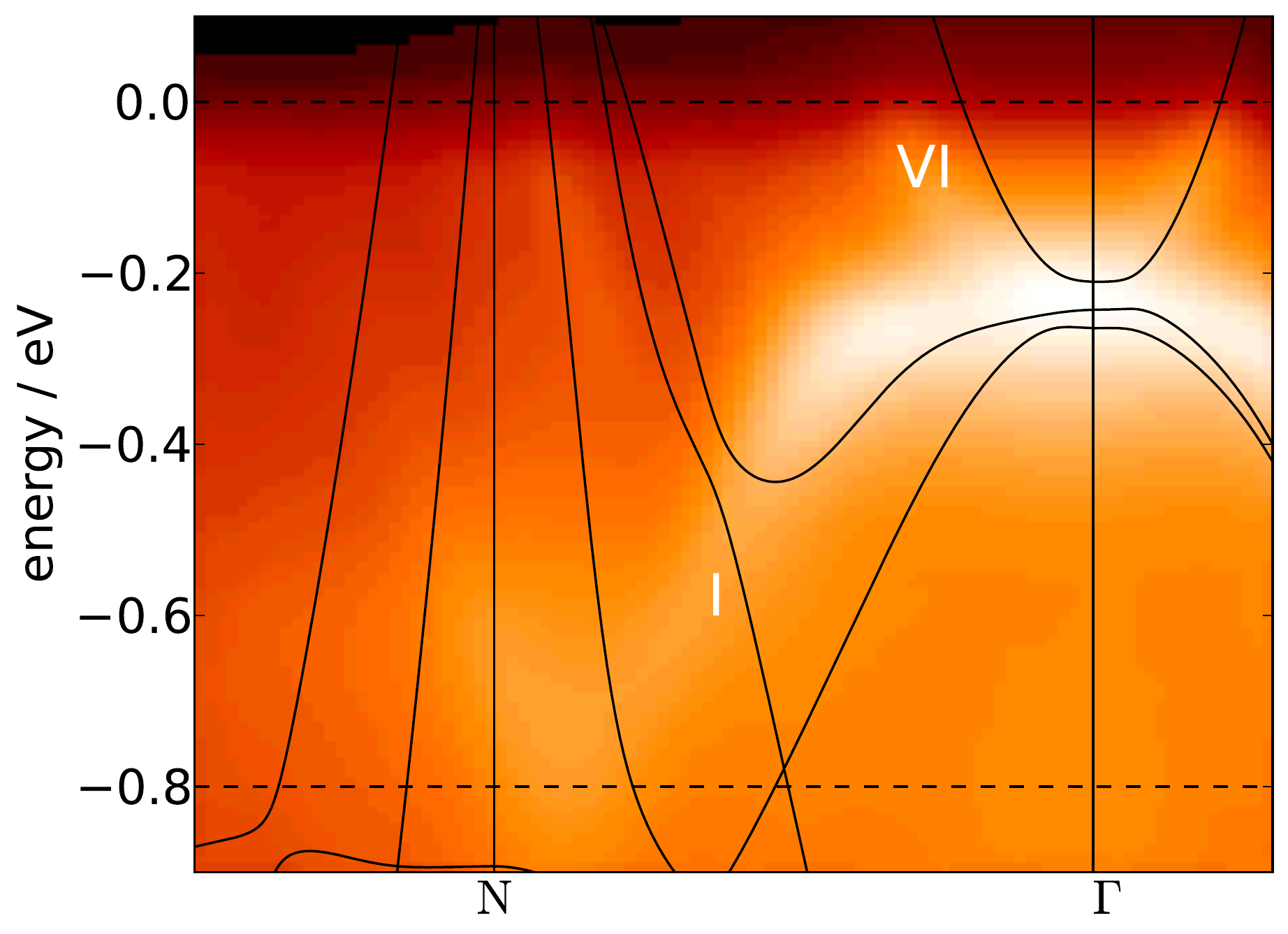}
\caption{Overlay of quasi-particle bands from LDA+Gutzwil\-ler
for $U=9.0\,{\rm eV}$, $J=0.54\,{\rm eV}$, and 
lattice parameter $a=5.39a_{\rm B}$ 
with ARPES data
along the N--$\Gamma$ direction.~\protect\cite{PhysRevB.72.155115}
\label{fig:soc_NG}}
\end{figure}

\begin{figure}[hb]
\includegraphics[width=0.45\textwidth]{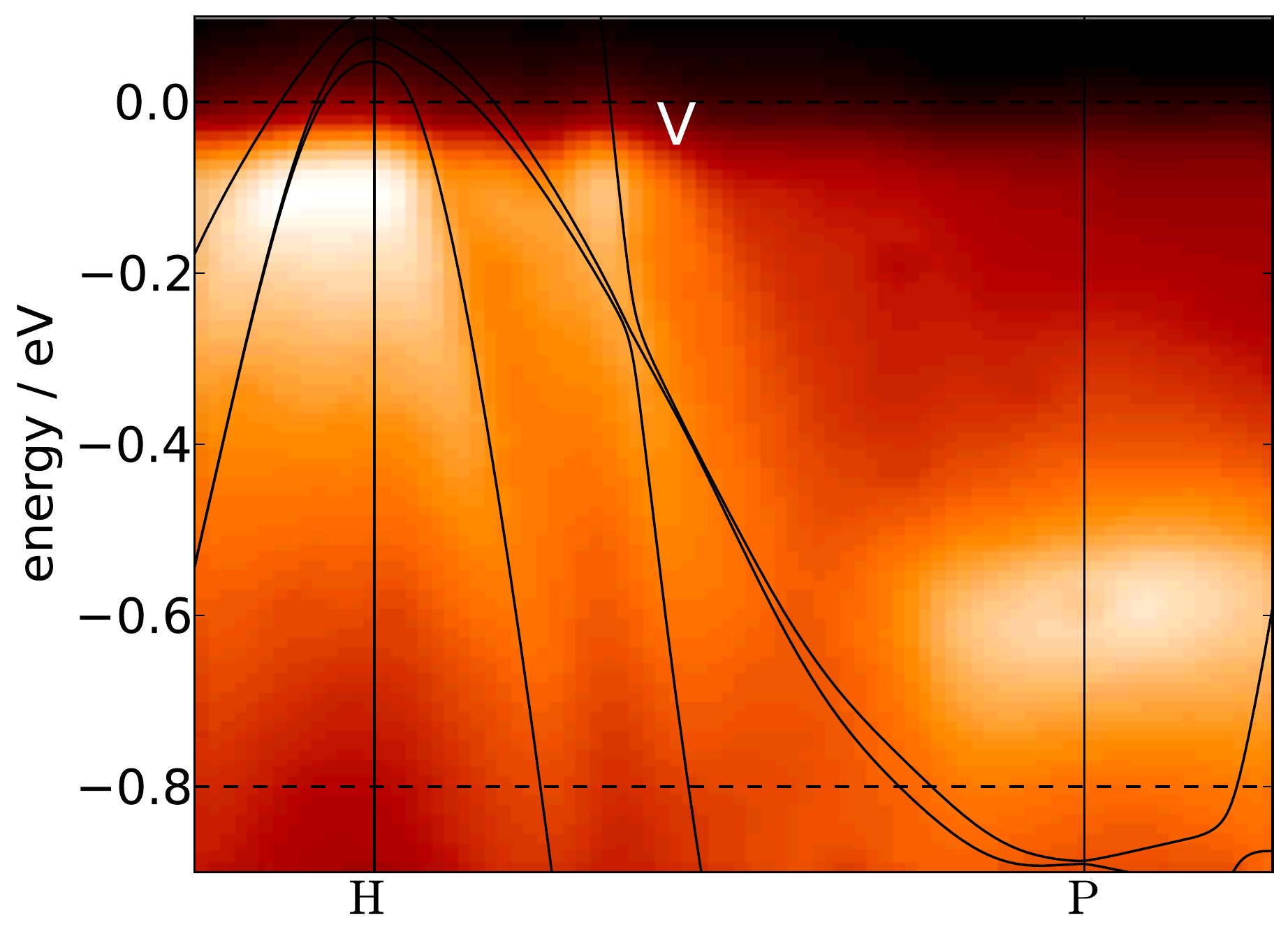}
\caption{Overlay of quasi-particle bands from LDA+Gutzwil\-ler
for $U=9.0\,{\rm eV}$, $J=0.54\,{\rm eV}$, and 
lattice parameter $a=5.39a_{\rm B}$ 
with ARPES data
along the H--P direction.~\protect\cite{PhysRevB.72.155115}
\label{fig:soc_HP}}
\end{figure}

Figures~\ref{fig:soc_GP} to~\ref{fig:soc_HP}
show an overlay of ARPES data from 
Sch\"afer~et~al.~\cite{PhysRevB.72.155115} with the results of our 
perturbative spin-orbit calculation based on the Gutzwiller-DFT.
A quantitative comparison 
between theory and experiment is given in table~\ref{tble:comp}
where we list the Fermi wavenumbers and velocities
for various directions and Fermi sheets 
LDA+Gutzwiller, and ARPES.~\cite{PhysRevB.72.155115}
For completeness, we also include in the table values from our fully relativistic
GGA calculations, see Sect.~\ref{ffd} for details.

We begin our discussion with the $\Gamma$--P high symmetry line 
in the Brillouin zone. From Fig.~\ref{fig:soc_GP}
we see that, close to the $\Gamma$-point,
we observe a very good agreement between
LDA+Gutzwiller and experimental data 
for the Fermi sheet~VI. 
Moreover, the Fermi wavenumbers and the velocities 
agree very well, as seen from table~\ref{tble:comp}.
Our LDA+Gutzwiller improves the theoretical values 
for the Fermi velocity. The
mass ratio between theory and experiment reduces from
$v_{\rm D}/v_{\rm R}=1.9$ in DFT(GGA) to $v_{\rm G}/v_{\rm R}=1.6$
in LDA+Gutzwiller.
Note that the mass ratio is unity for a perfect agreement between theory
and experiment, and it is larger than unity when the theoretical mass
is smaller than measured value, $v/v_{\rm R}=m_{\rm R}/m$.

The largest discrepancies between 
experiment and LDA+Gutzwiller theory are seen at and around the P-point. 
The LDA+Gutzwiller bands are about $0.3\,{\rm eV}$ below the ARPES bands,
and the discrepancy in LDA+ Gutzwiller is actually {\sl worse\/}
than in LDA(GGA). We do not have an explanation for this deviation.

Half way on the line $\Gamma$--P 
there is the Fermi sheet~I. For this band, the values for the Fermi wavenumbers
from DFT(GGA) and LDA+Gutzwiller are very close to the experimental value but
the Fermi velocities deviate considerably, even though
LDA+Gutzwiller has a slightly better mass ratio, 
$v_{\rm G}/v_{\rm R}=2.8$ versus $v_{\rm D}/v_{\rm R}=3.4$.

In Fig.~\ref{fig:soc_GH} we plot the data overlay
along the high-symmetry line $\Gamma$--H.
As discussed before, the agreement of the quasi-particle bands, 
wavenumbers and velocities of Fermi sheet~VI close 
to the $\Gamma$-point is very good.
The same holds true for the Fermi sheet~II close to the
H-point. 

For the majority Fermi sheet~I half way between
the points~$\Gamma$ and~H we find 
a large mass ratio as in LDA(GGA),
$v_{\rm G}/v_{\rm R}=3.0$ versus $v_{\rm D}/v_{\rm R}=2.1$.
Note, however, that several bands meet at the Fermi energy 
with the same Fermi wavenumber, and the spin-orbit coupling leads
to a splitting of bands. Therefore, 
it is difficult to determine the Fermi velocity
due to the sequence of crossings.
This region around the Fermi energy is not very suitable for
a meaningful comparison between theory and experiment.

In Fig.~\ref{fig:soc_NG} we plot the data overlay 
along the high-symmetry line N--$\Gamma$.
In experiment, the intensity along this direction is suppressed close to
the point~N due to matrix-element effects. However, we reproduce
a crossing of the Fermi energy close to the N-point 
and the slope of the bands agree quite well. 

For the minority band~VI close to the 
$\Gamma$-point we find a mass ratio of 
$v_{\rm G}/v_{\rm R}=1.6$, versus $v_{\rm D}/v_{\rm R}=1.9$ in DFT(GGA).
For the majority band~I half way between $\Gamma$ and~N we find a
mass ratio of
$v_{\rm G}/v_{\rm R}=1.3$, versus $v_{\rm D}/v_{\rm R}=1.6$ in DFT(GGA). 
In both cases, we find a slight improvement over the DFT(GGA) results.
Half way between the points~N and $\Gamma$,
it seems as if the LDA+Gutzwiller bands 
at energies of about $-0.5\,\mathrm{eV}$
do not agree very well with the 
ARPES bands but better experimental data are needed 
for a definitive statement.

Lastly, we take a look at the direction H--P
in Fig.~\ref{fig:soc_HP}.
As already seen from the other plots, 
the agreement at and around the point~H is quite good while
the comparison at point~P reveals some discrepancies between
theory and experiment. In addition, the ARPES data show some 
distinct Fermi level crossing half way between~H and~P. 
The LDA+Gutzwiller approach yields a Fermi wavenumber 
for this crossing that deviates slightly from experiment 
with a mass ratio
$v_{\rm G}/v_{\rm R}=2.3$ versus $v_{\rm D}/v_{\rm R}=2.7$ in DFT(GGA). 

Depending on the Fermi wavevector,
the quasi-par\-ticle mass in Gutzwiller-DFT is some 20\%
larger than in DFT(GGA). A similar mass enhancement is observed
in DFT(DMFT) calculations.~\cite{PhysRevB.90.155120,PhysRevLett.103.267203}
However, as seen from table~\ref{tble:comp},
the mass ratio between theory and experiment is consistently larger than unity,
$v_{\rm G}/v_{\rm R}=m_{\rm R}/m_{\rm G}>1$, and strongly depends on the 
Fermi wavevector.
The correlation-induced mass enhancement alone cannot account for the large
mass renormalization as seen in experiment.
The resolution of this discrepancy remains one of the incompletely understood
problems for iron and other magnetic materials.

\section{Conclusions}
\label{sec:conclusions}

In this work, we used the Gutzwiller-DFT for a detailed study 
of the ground-state properties and the quasi-particle bandstructure of iron. 
We find that, 
for a Hubbard interaction of $U=9\, {\rm eV}$ and a Hund's-rule coupling of
$J=0.54\, {\rm eV}$, we 
reproduce the experimental lattice parameter and magnetization, and we
obtain the bulk modulus of ferromagnetic bcc iron
in very good agreement with experiment.
Upon increasing pressure we qualitatively
reproduce the transition to non-magnetic hcp iron.

We find that the ground-state magnetization sensitively depends on
the Hund's-rule coupling~$J$. In contrast to physical intuition,
an increase of~$J$ leads to a decrease of the magnetization.
For example, at $U=9\, {\rm eV}$, an increase from $J=0.45\, {\rm eV}$ 
to $J=0.68\, {\rm eV}$ decreases the magnetization from $m=2.5\mu_{\rm B}$
to $m=2.05\mu_{\rm B}$. The Hund's-rule coupling generates
a splitting of iso-electronic atomic levels, and the corresponding 
redistribution of local occupancies considerably impedes the electrons' motion
through the lattice (`configurational hopping blockade'). 
In a band magnet, the delicate balance 
between the Hund's-rule and exchange-energy gains
against the corresponding losses in kinetic energy
makes the magnetization sensitive to the Hund's-rule coupling.
Therefore, the absolute value of~$J$ is much more decisive 
for physical quantities 
than the value of the Hubbard interaction.
For the calculation of some physical quantities,
a larger value of~$J$ can be `traded in' for a smaller~$U$.

Gutzwiller-DFT renormalizes the quasi-particle bands as obtained from DFT(LDA).
While the $4sp$-type parts are almost unchanged,
the $3d$-bands are shifted and their width is reduced, in agreement with experiment.
Shifts and renormalizations are also observed when we compare 
the Gutzwiller-DFT results with GGA calculations
although the effects are quantitatively smaller.
The applied double-counting corrections make sure that the
average $3d$-electron density remains 
essentially the same, $\overline{n}_d\approx 7.2$.
The agreement between the Gutzwiller quasi-particle bands with
ARPES data is fairly good when we take spin-orbit effects into account perturbatively.
In general, Gutzwiller-DFT agrees better with ARPES data than
DFT, 
both in the LDA and GGA approximations.

Finally, 
we note that the optimal atomic parameters in the present Gutzwil\-ler-DFT study 
on iron resemble those used by Deng~et alii,~\cite{Deng2008, Deng2009}
but are sensibly higher than those used in  a more recent Gutzwiller-DFT work 
by Borghi et al.~\cite{PhysRevB.90.125102}
who propound a $U$-para\-me\-ter of $U_{\rm B}=2.5\, {\rm eV}$ for iron that 
is significantly smaller than our values.
Part of the discrepancy is probably due to their different choice of energy 
window and basis set for the construction of the many-body model.
However, there are also substantial differences already 
at  the bare DFT(LDA) level ($U=J=0$); Borghi et al.\ use 
a localized orbital code (SIESTA) whose results deviate from
other DFT codes for iron.
Table~III of Ref.~[\onlinecite{PhysRevB.90.125102}] gives 
the lattice constant $a_{\rm Borghi}=2.83\, \hbox{\AA}=5.35 a_{\rm B}$,
while we find $a_0=5.2 a_{\rm B}$, in agreement 
with earlier LDA-LAPW calculations.~\cite{PhysRevLett.54.1852}
Since the lattice constant monotonically increases as a function of~$U$, 
see Fig.~\ref{fig:lat-const}, 
it is not surprising that Borghi et al.\ require a smaller Hubbard interaction to 
reproduce the experimental lattice parameter $a=5.42a_{\rm B}$,
and claim a different role of electronic correlations in iron.

Despite the improvements of Gutzwiller-DFT and DFT(DMFT) 
over standard DFT(LDA 
and GGA), 
the theoretical Fermi velocities are typically too large, i.e.,
the quasi-particle masses from theory are too low in comparison with experiment.
This systematic discrepancy could have several reasons.
First, it might be necessary to process the theoretical bandstructures 
further to mimic the excitation process 
in ARPES experiments.~\cite{1367-2630-12-1-013007}
However, this approach could not explain why the systematic
mass enhancement is also seen in de-Haas--van-Alphen 
measurements for iron  and ferromagnetic nickel-compounds.~\cite{JMMM-Lonzarich}
Therefore, it is more likely that the effective mass results from the
interaction of the quasi-particles with low-energy magnetic excitations, i.e.,
magnetic polarons exist near the Fermi energy.~\cite{PhysRevLett.92.097205}
At present, however, the inclusion of long wave-length excitations is beyond
the Gutzwiller-DFT.

\begin{acknowledgments}
We thank R.~Claessen for helpful discussions on the mass enhancement in
iron, and him and J.~Sch\"afer for sending us 
the figures of their experimental data.
L.B.\ would like to thank M.~Aichhorn for useful discussion 
and for pointing out reference~[\onlinecite{PhysRevB.90.155120}].

The work was supported in part by 
the SPP~1458 of the Deutsche Forschungsgemeinschaft (BO~3536/2 and GE~746/10).
The figures in this publication were created using the matplotlib
library.~\cite{Hunter:2007}
\end{acknowledgments}

\appendix

\section{Atomic interactions and double counting}
\label{app:Coulomb}

For the $3d$ shell of $e_g$ and $t_{2g}$  orbitals in transition metals
the local Hamiltonian~(\ref{eq:VlocR}) reads ($\sigma=\uparrow,\downarrow$)
\begin{eqnarray}\nonumber
&&\hat{V}_{\rm loc}=
\frac{1}{2}
\sum_{c,\sigma}U(c,c)
\hat{n}_{c,\sigma}\hat{n}_{c^{\prime},\bar{\sigma}}
+\frac{1}{2}\!\!
\sum_{\genfrac{}{}{0pt}{2}{c(\neq)c^{\prime}}{\sigma,\sigma^{\prime}}}
\!\!\widetilde{U}_{\sigma,\sigma^{\prime}}(c,c^{\prime})
\hat{n}_{c,\sigma}\hat{n}_{c^{\prime},\sigma^{\prime}} \\
\nonumber 
&&+
\frac{1}{2}
\sum_{c(\neq)c^{\prime}}J(c,c^{\prime})\left(\hat{c}^{\dagger}_{c,\uparrow}\hat{c}^{\dagger}_{c,\downarrow}
\hat{c}_{c^{\prime},\downarrow}\hat{c}_{c^{\prime},\uparrow}+ {\rm h.c.}\right) 
 \\
&&+\frac{1}{2}
\sum_{c(\neq)c^{\prime};\sigma}J(c,c^{\prime})\hat{c}^{\dagger}_{c,\sigma}
\hat{c}^{\dagger}_{c^{\prime},\bar{\sigma}}
\hat{c}_{c,\bar{\sigma}}\hat{c}_{c^{\prime},\sigma}\nonumber\\
&&+\bigg[\sum_{t; \sigma,\sigma^{\prime}}
(T(t)-\delta_{\sigma,\sigma^{\prime}}A(t))
\hat{n}_{t,\sigma}\hat{c}^{\dagger}_{u,\sigma^{\prime}}\hat{c}_{v,\sigma^{\prime}}\nonumber\\
&&\hphantom{+\bigg[}
+\sum_{t,\sigma}A(t)
\left(
\hat{c}^{\dagger}_{t,\sigma}\hat{c}^{\dagger}_{t,\bar{\sigma}}
\hat{c}_{u,\bar{\sigma}}\hat{c}_{v,\sigma}+
\hat{c}^{\dagger}_{t,\sigma}\hat{c}^{\dagger}_{u,\bar{\sigma}}
\hat{c}_{t,\bar{\sigma}}\hat{c}_{v,\sigma}
\right)\nonumber \\\nonumber
&&\hphantom{+\bigg[}
+\sum_{ 
\genfrac{}{}{0pt}{2}{t(\neq)t^{\prime}(\neq)t^{\prime \prime}}{e,\sigma,\sigma^{\prime}}}
S(t,t^{\prime};t^{\prime \prime},e)
\hat{c}^{\dagger}_{t,\sigma}\hat{c}^{\dagger}_{t^{\prime},\sigma^{\prime}}
\hat{c}_{t^{\prime \prime},\sigma^{\prime}}\hat{c}_{e,\sigma}+{\rm h.c.}\bigg ]\; .
\\
\label{app3.5}
\end{eqnarray}
Note that the factors $1/2$ in the first three lines have been erroneously
missing in our previous 
publications~[\onlinecite{1367-2630-16-9-093034}] 
and [\onlinecite{Schickling2011}].
Here, $\widetilde{U}_{\sigma,\sigma^{\prime}}(c,c^{\prime})=
U(c,c')-\delta_{\sigma,\sigma'}J(c,c')$, and
we suppressed the site index~$\vecR$. The index~$c$ 
sums over all five $d$-orbitals 
while $t$ and $e$ are indices for the  
three $t_{2g}$ orbitals with symmetries $xy$, $xz$, and $yz$
and the two $e_g$ orbitals with symmetries $u=3z^2-r^2$ and $v=x^2-y^2$, 
respectively. 
Of all the parameters $U(c,c^{\prime})$, $J(c,c^{\prime})$,
$A(t)$, $T(t)$, $S(t,t^{\prime};t^{\prime \prime},e)$ 
only ten are independent in cubic 
symmetry.~\cite{1367-2630-16-9-093034,Sugano1970}
When we assume that all 3$d$-orbitals have 
the same radial wavefunction (`spherical approximation'), 
all parameters are determined by, e.g., the 
three Racah parameters $A,B,C$. 
They are related to the Slater-Condon parameters via
\begin{equation}
A=F^{(0)}-\frac{F^{(4)}}{9}
\;, \; B=\frac{1}{49}\Bigl(F^{(2)}-\frac{5}{9}F^{(4)}\Bigr)
\;, \; C=\frac{5}{63}F^{(4)}\; ,
\end{equation}
or, inversely,
\begin{equation}
F^{(0)}=A+\frac{7}{5}C
\quad, \quad F^{(2)}=49 B +7C 
\quad, \quad F^{(4)}=\frac{63}{5}C\; .
\end{equation}
Explicit expressions for the relations between the parameters in eq.~(\ref{app3.5})
and the Racah parameters~$A,B,C$ can be found in appendix~C of 
Ref.~[\onlinecite{1367-2630-16-9-093034}].
For comparison with other work, we 
introduce the Coulomb interaction between electrons in the
same 3$d$-orbitals (intra-orbital Hubbard interaction),
\begin{equation}
U=A+4B+3C\; ,
\end{equation}
the average Coulomb interaction 
between electrons in different orbitals (inter-orbital Hubbard interaction),
\begin{equation}
U^{\prime}=\frac{1}{10}\sum_{c<c^{\prime}_{\displaystyle\vphantom{X}}} 
U(c,c^ {\prime})=A-B+C\; ,
\end{equation}
and the average Hund's-rule exchange interaction,
\begin{equation}
J=\frac{1}{10}\sum_{c<c'}J(c,c')=\frac{5}{2}B+C\; . 
\label{eq:appJdef}
\end{equation}
These three quantities are not independent but related by
the symmetry relation $U'=U-2J$.
This means that by choosing two of these parameters (e.g., $U$ and 
 $J$) the three Racah parameters, and therefore all the parameters 
 in~Eq.~(\ref{app3.5}) are not uniquely defined. Hence, we use the additional 
 relation $C/B=4$ which is a reasonable assumption for transition 
 metals.~\cite{Sugano1970}
It corresponds to $F^{(2)}/F^{(4)}=55/36=1.53$, in agreement with
the estimate $F^{(2)}/F^{(4)}\approx 1.60=8/5$ 
by de Groot et alii.~\cite{PhysRevB.42.5459}
For completeness, we give the dependencies of the Racah parameters on
$U$ and $J$ 
\begin{equation}
A=U-\frac{32}{13}J \quad , \quad B=\frac{2}{13} J\quad , \quad C=\frac{8}{13} J\; .
\end{equation}
In our previous study on nickel,\cite{1367-2630-16-9-093034}
we have tested three different 
types of double-counting corrections that had been proposed 
in the literature. It turned out that only one of them 
leads to sensible results for nickel. As for nickel~\cite{1367-2630-16-9-093034}
we employ the widely used functional~\cite{ PhysRevB.44.943,PhysRevB.52.R5467} 
\begin{equation}
\label{78}
V_{\rm dc}=\frac{\bar{U}}{2}\bar{n}(\bar{n}-1)-
\frac{\bar{J}}{2}\sum_{\sigma}\bar{n}_{\sigma}(1-\bar{n}_{\sigma}) \;.
\end{equation}
For $3d$-electrons we have
\begin{equation}
F^{(0)}=\bar{U}=\frac{1}{5}\left(U+4U^{\prime}\right) \quad ,\quad
\bar{J}=\bar{U}-U^{\prime}+J \;, 
\label{eqApp:barU}
\end{equation}
and
\begin{equation}
\bar{n}_{\sigma}\equiv \sum_{c=1}^{N_c} C_{c,c;\sigma}^{\rm G}\quad, \quad 
\bar{n}\equiv\bar{n}_{\uparrow}+\bar{n}_{\downarrow} \; ,
\end{equation}
where $N_c$ is the number of correlated orbitals, in our case, $N_c=5$.
Moreover, 
\begin{equation}
C_{c,c;\sigma}^{\rm G}=\frac{\langle \Psi_{\rm G} | 
\hat{c}_{\vecR,c,\sigma}^+\hat{c}_{\vecR,c,\sigma}^{\vphantom{+}}
| \Psi_{\rm G}\rangle}{\langle \Psi_{\rm G} | \Psi_{\rm G}\rangle}
=\langle \psi_0 | \hat{c}_{\vecR,c,\sigma}^+\hat{c}_{\vecR,c,\sigma}^{\vphantom{+}}
| \psi_0\rangle 
\end{equation}
is the $\sigma$-electron density for the correlated $3d$ orbital~$c$
in the Gutzwiller wavefunction. Note that the second equality only holds
in the limit of infinite dimensions for our $e_g$-$t_{2g}$~orbital 
structure.~\cite{1367-2630-16-9-093034}
   
%

\end{document}